\documentclass[aps,pre,10pt,twocolumn,superscriptaddress,balancelastpage,showpacs,letterpaper,lengthcheck]{revtex4-1}
\usepackage{centernot}
\usepackage{graphicx}
\usepackage{amsmath}
\usepackage{times}
\usepackage{amssymb}
\usepackage{mathrsfs}
\usepackage{chemarr}
\usepackage{color}
\usepackage{url}
\usepackage{version}
\usepackage[ pdftex,colorlinks=true,
pdfstartview=FitV,
linkcolor= linkcolor,
citecolor= linkcolor,
urlcolor= linkcolor,
hyperindex=true,
hyperfigures=false]
{hyperref}
\definecolor{linkcolor}{rgb}{0,0,0.6} 
\definecolor{brown}{RGB}{112,10,10}

\graphicspath{{figures},{figures/eps/},{figures/pdf/}}

\newcommand{\dd}{\text{d}}
\newcommand{\ee}{\text{e}}
\newcommand{\ii}{\text{i}}
\newcommand{\eps}{\varepsilon}
\newcommand{\va}{\xi_\text{\tiny NG}}
\newcommand{\cO}{\mathcal{O}}
\newcommand{\td}{\tau_\text{\tiny R}}
\newcommand{\Dt}{D}
\newcommand{\Da}{D_\text{\tiny A}}
\newcommand{\da}{\varepsilon}
\newcommand{\sig}{\sigma}
\newcommand{\ft}{f_t}
\newcommand{\tp}{t^*}
\newcommand{\tw}{t_\text{i}}
\newcommand{\ta}{\tp}
\newcommand{\p}{\partial}
\newcommand{\cW}{\mathcal{W}}
\newcommand{\Fs}{F_\text{\tiny S}}
\newcommand{\Gs}{G_\text{\tiny S}}

\providecommand{\avg}[1]{\left \langle #1 \right \rangle}
\providecommand{\avgc}[1]{\left \langle #1 \right \rangle_\text{\tiny C}}
\providecommand{\pnt}[1]{\left  ( #1 \right ) }
\providecommand{\brt}[1]{\left  [ #1 \right ] }
\providecommand{\cur}[1]{\left \{ #1 \right \}}
\providecommand{\abs}[1]{\left | #1 \right|}
\providecommand{\f}[2]{\frac{ #1}{#2}}
\providecommand{\df}[2]{\dfrac{ #1}{#2}}

\begin{document}


\title{Active cage model of glassy dynamics}

\author{\'{E}tienne Fodor}
\affiliation{Laboratoire Mati\`ere et Syst\`emes Complexes, UMR  7057 CNRS/P7, Universit\'e Paris Diderot, 10 rue Alice Domon et L\'eonie Duquet, 75205 Paris cedex 13, France}

\author{Hisao Hayakawa}
\affiliation{Yukawa Institute for Theoretical Physics, Kyoto University, Kitashirakawa-oiwake cho, Sakyo-ku, Kyoto 606-8502, Japan}

\author{Paolo Visco}
\affiliation{Laboratoire Mati\`ere et Syst\`emes Complexes, UMR  7057 CNRS/P7, Universit\'e Paris Diderot, 10 rue Alice Domon et L\'eonie Duquet, 75205 Paris cedex 13, France}

\author{Fr\'ed\'eric van Wijland}
\affiliation{Laboratoire Mati\`ere et Syst\`emes Complexes, UMR  7057 CNRS/P7, Universit\'e Paris Diderot, 10 rue Alice Domon et L\'eonie Duquet, 75205 Paris cedex 13, France}
\affiliation{Yukawa Institute for Theoretical Physics, Kyoto University, Kitashirakawa-oiwake cho, Sakyo-ku, Kyoto 606-8502, Japan}


\date{\today} \pacs{}

\begin{abstract}
  We build up a phenomenological picture in terms of the effective dynamics of a tracer confined in a cage experiencing random hops to capture some characteristics of glassy systems. This minimal description exhibits scale invariance properties for the small-displacement distribution that echo experimental observations. We predict the existence of exponential tails as a cross-over between two Gaussian regimes. Moreover, we demonstrate that the onset of glassy behavior is controlled only by two dimensionless numbers: the number of hops occurring during the relaxation of the particle within a local cage, and the ratio of the hopping length to the cage size.

\end{abstract}

\maketitle


\section{Introduction}

Time-dependent density correlations in an atomic or colloidal glass former exhibit, as temperature is decreased (or as density is increased) a two step relaxation. The short time $\beta$ relaxation is associated with localized motion of the particles while the longer time $\alpha$ relaxation is associated with cooperatively rearranging regions (CRR). This behavior of the dynamic structure factor is well accounted for~\cite{Biroli:11}, at sufficiently high temperatures, by the Mode-Coupling Theory (MCT)~\cite{Gotze}. MCT being an approximate theory it does have its share of pitfalls but these are well-documented, and it is fair to say that they are not fully understood, let alone on intuitive grounds. One such predictions is the existence of a critical temperature below which the $\alpha$-relaxation stage extends to arbitrary large times. This is often connected to the difficulty of capturing the dynamics of the CRR's, which is built from rare and intermittent events. One recent research direction has been to brute force improve the theoretical basis of the MCT, and recent promising results~\cite{PhysRevLett.115.205701} seem to indicate that this may pay off to eliminate some of its unpleasant features. Regarding the dynamics (diffusive vs. arrested) of an individual tracer~\cite{2015arXiv151100254P,2015arXiv151100786B} similar efforts are being made at looking beyond the existing MCT. The most recent experiments, instead of focussing on  the relaxation of collective density modes, are based on the observation of individual tracers. Tracking experiments have been  conducted by many different groups, both in actual colloidal systems~\cite{Weeks:00,Kegel:00,Weeks:02,Kilfoil:07,Kilfoil:09,Gardel:13} as well as in silico~\cite{Stariolo:06,Berthier:07}. The increasing accuracy of these methods allows one to measure the probability distribution function of the tracer displacement with unprecedented statistics~\cite{Gao:09}.

To analyze and interpret tracking experiments, minimal phenomenological models have been developed. By using a continuous time random walk (CTRW) description, glassy systems have been studied numerically~\cite{Farago:14a,Farago:14b}, and exact analytic results have been obtained for the Van Hove correlation function~\cite{Berthier:07}. The CTRW considers instantaneous transitions between locally stable positions, thus decoupling the diffusion in a confined environment from the cage jumps~\cite{Chandler:07}. Some model variations include the existence of a multiplicity of time scales in the cage dynamics~\cite{Pastore:14,Pastore:14b,Pastore:15,Helfferich}, whereas others which include a single waiting time scale successfully reproduce many features of the dynamics of tracers in glassy systems~\cite{Berthier:07,Berthier:08}. Despite the success of these minimal models, the existence of a scale invariant regime for the displacement distribution, as observed experimentally~\cite{Weeks:00,Weeks:02,Kilfoil:09}, is still an open issue. The main purpose of this work is to show that this scale invariance feature can emerge from a phenomenological description.

Note that systems other than glasses fall within the scope of our study, such as sheared fluids~\cite{Berthier:00, Barrat:02, Berthier:02, Berthier:03}, or interacting self-propelled particles~\cite{Stuart:13, Berthier:14, Levis:14, Rao:14, Gomper:14}. For these systems glassy behavior is generally investigated through the self-intermediate scattering function (sISF). In the case of self-propelled particles it has been observed that the onset of glassy behavior is progressively shifted when self-propulsion increases. Even though a direct extension of MCT is enough to account for this phenomenon~\cite{Kurchan:13, Szamel:15, Brader:14}, an effective one-body dynamics that clearly determines how self propulsion affects the glassy behavior is still missing. We present here a minimal model that describes such one-body dynamics. By investigating the displacement distribution, we show that it displays scale invariance for small displacement, and we  determine how the onset of glassy behavior is linked to self propulsion. Finally, we introduce an effective mean--field potential to bridge over back to a system of interacting particles.


\begin{figure*}
\centering
\includegraphics[width=2\columnwidth]{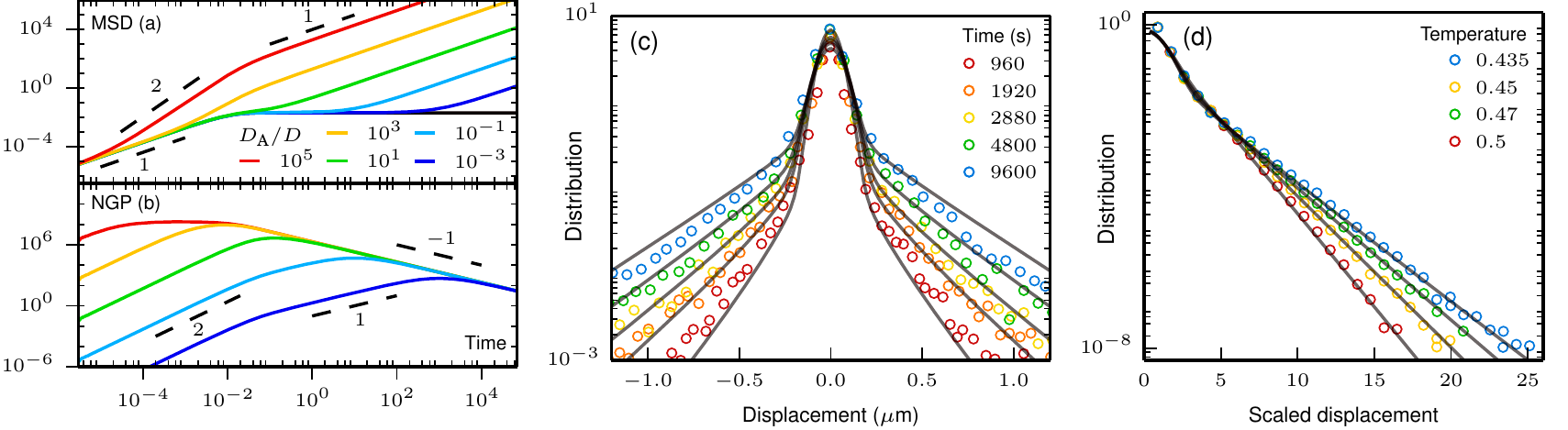}
\caption{\label{fig:panels}
Time-evolution of~(a) the mean-square displacement (MSD), and~(b) the non-Gaussian parameter (NGP). The MSD saturates for a steady cage (solid black line), corresponding to a vanishing NGP.
(c)~Comparison of the theoretical prediction of the distribution of displacement with that of  colloidal particles at density $\varphi = 0.429$ taken from \cite{Kilfoil:09}.
(d)~Theoretical distribution of displacement scaled by the standard deviation of the central Gaussian part corresponding to a binary Lennard-Jones glass-forming mixture for different temperatures at the $\alpha$ relaxation time~\cite{Berthier:07b}.
See the details in Appendix \ref{app:dataanalysis}.
}
\end{figure*}

\section{Model}
Our model is based on a common picture which has emerged for glassy and granular systems connecting the slowing down of the dynamics to the cage effect~\cite{Berthier:07,Pastore:14,Pastore:15,Scalliet:15,Lasanta:15}. The escape of the particle from the local confinement stems from structural rearrangements of the system. The activation of such sudden and irreversible reorganization can be of three types: spontaneous thermal fluctuations, external shear, or internal self-propulsion. In what follows, we refer to such a process as a directed event, by contrast to the passive diffusion of the particle within the cage, whose precise structure is given by collective many--body effects of the global system.

We consider a particle confined within a harmonic cage of typical size $\sig$. We introduce the time scale $\td$ quantifying the time needed by the particle to explore the cage. The fluctuations of the particle in the cage are driven by a noise of amplitude $D=\sig^2/\td$. To account for structural rearrangements, we assume that the central position of the cage is subjected to random shifts. This is to mimic the modification of the metastable state explored by the particle. The cage hops instantaneously by a random distance, which is exponentially distributed with a characteristic hopping length $\da$. The time between two consecutive hops is also exponentially distributed, with a mean value $\tau_0$. After a cage rearrangement, the particle relaxes towards the new cage position. We regard such relaxation as an equilibrium process. The fluctuation-dissipation theorem enforces that the relaxation time should equal the typical time of exploration  $\td$. We end up describing the one-dimensional dynamics of the particle position $x$ as
\begin{equation}\label{eq:dyn}
\f{\dd x}{\dd t} = -\f{x(t)-x_0(t)}{\td} + \xi_\text{\tiny G}(t), \quad \f{\dd x_0}{\dd t} = \va(t),
\end{equation}
where $\xi_\text{\tiny G}$ is a zero-mean Gaussian white noise with correlations $\avg{\xi_\text{\tiny G}(t)\xi_\text{\tiny G}(t')}=2\Dt\delta(t-t')$. More realistic higher-dimensional generalizations do not induce any physical difference with the one-dimensional modeling we adopt here.

Our motivations for the explicit form of the noise $\va$ acting on the cage are twofold. One the one hand, we expect the rare intermittent events behind the $\alpha$ relaxation to be unable to build up uncorrelated Gaussian statistics. On the other hand, we choose a specific form which has the advantage of allowing for a closed form analytic solution that will ease the subsequent analysis. It leads us to consider a zero-mean non-Gaussian white noise with cumulants
\begin{equation}\label{eq:va}
\avgc{\va(t_1)\dots \va(t_{2n})}=(2n)!\f{\da^{2n}}{\tau_0}\delta_{2n}(t_1,\dots, t_{2n}).
\end{equation}
We decouple caging and hopping dynamics, so that $\xi_\text{\tiny G}$ and $\va$ are uncorrelated processes. For symmetry reasons, only the $2n$-time correlation functions of $\va$ are non zero. Additional arguments on the robustness of our results with a generic time symmetric non-Gaussian white noise is given in Appendix~\ref{app:nongaussian}. We will explicitly demonstrate that the scale invariance of the small displacement distribution is indeed insensitive to the specifics of the hop distribution, but the exponential crossover regime to a Gaussian rests on a typical exponential distribution of cage hops.

We distinguish the passive and active fluctuations of the particle. The former are associated with the confined motion of the particle in a steady cage, as measured by $D$. The latter are induced by the cage hops, thus describing the motion in a non-confined environment characterized by the diffusion coefficient $\Da=\da^2/\tau_0$. The coexistence of both Gaussian and non-Gaussian noises is crucially important to enhance non-Gaussian nature~\cite{Kanazawa:15}.

\section{Statistics of displacement}

\subsection{Fourier distribution of displacement}

We are interested in the fluctuations of the displacement $\Delta x(t,\tw) = x(t+\tw)-x(\tw)$, which no longer depends on the initial time of measurement $\tw$ in the limit $\tw\to\infty$. We define the probability distribution of displacement in the Fourier domain as
\begin{equation}\label{eq:defP}
\tilde P(q,t) = \underset{\tw\to\infty}{\lim} \avg{\ee^{\ii q \Delta x(t,\tw)} }.
\end{equation}
Since the passive and active processes are uncorrelated, we separate the displacement distribution as $\tilde P=\tilde P_\text{\tiny P}\tilde P_\text{\tiny A}$, where the subscripts P and A refer respectively to passive and active contributions. The passive distribution is Gaussian, thus being entirely determined by the passive mean-square displacement
\begin{equation}\label{eq:Ptt}
\tilde P_\text{\tiny P}(q,t) = \ee^{-(q\sig)^2\ft},
\end{equation}
where $\ft=1-\ee^{-t/\td}$. We express the active distribution as
\begin{equation}\label{eq:PA}
\tilde P_\text{\tiny A}(q,t) = \underset{\tw\to\infty}{\lim} \avg{\exp\brt{\ii q \int_0^\infty \dd s h(s) \va(s)}},
\end{equation}
where
\begin{eqnarray}\label{eq:h}
h(s) &=& \brt{ 1 - \ee^{-(\tw+t-s)/\td} } \Theta(\tw+t-s)\Theta(s-\tw)
\nonumber
\\
& & + \ee^{-(\tw-s)/\td} \brt{ 1 - \ee^{-t/\td} } \Theta(\tw-s) ,
\end{eqnarray}
with $\Theta$ being the Heaviside step function. We evaluate the average in Eq.~\eqref{eq:PA} with the expression of the characteristic functional of the white non-Gaussian noise $\va$~\cite{Budini,Baule:09R,Baule:09}
\begin{eqnarray}
\tilde P_\text{\tiny A}(q,t) &=& \underset{\tw\to\infty}{\lim}\exp\brt{\sum_{n=1}^\infty\f{(\ii q\da)^{2n}}{\tau_0} \int_0^\infty \dd s \brt{h(s)}^{2n}}
\\
\label{eq:Pint}
&=& \underset{\tw\to\infty}{\lim}\exp\brt{-\f{1}{\tau_0}\int_0^\infty \dd s \f{\brt{q\da h(s)}^2}{1+\brt{q\da h(s)}^2}} ,
\end{eqnarray}
where we have used the explicit form of the noise cumulants in Eq.~(\ref{eq:va}). We compute the integral in Eq.~\eqref{eq:Pint} by using the explicit expression of $h$ in Eq.~\eqref{eq:h}, and we take the limit of large $\tw$ after the integration. From this, we arrive at the result:
\begin{eqnarray}\label{eq:tP}
\tilde P(q,t) &=& \exp \cur{ -\f{ \td q\da /\tau_0 }{1+(q\da)^2} \brt{ \f{q\da t}{\td}  -\arctan\pnt{q\da\ft} } }
\nonumber
\\
& &\times\brt{1+(q\da\ft)^2}^{-\f{\td}{2\tau_0}\f{(q\da)^2}{1+(q\da)^2}}\ee^{-(q\sig)^2\ft}.
\end{eqnarray}
This result contains all the statistical information related to the displacement $\Delta x(t)$. In particular, all the moments are defined as: 
\begin{equation}
\label{eq:moments}
\avg{\Delta x^n(t)} = \underset{\tw\to\infty}{\lim} \avg{ \Delta x^n(t,\tw) } = \left.\f{\p^n \tilde P}{\p (\ii q)^n}\right|_{q=0}.  
\end{equation}
\subsection{Mean-square displacement and non-Gaussianity}

As a first insight into the dynamics of our model, we study the time evolution of the second moment $\avg{\Delta x^2(t)}$, {\it i.e.}  the mean-square displacement (MSD). Its expression can be computed from (\ref{eq:moments}):
\begin{equation}
\text{MSD} = 2 \pnt{\Dt-\Da}\td \ft + 2 \Da t,
\end{equation}
The behavior of the MSD is controlled by three independent parameters $\{\Dt,\Da,\td\}$. It is diffusive at short and long times with diffusion coefficients $\Dt$ and $\Da$, respectively [Fig.~\ref{fig:panels}(a)]. The predictions for a steady and a hopping cage coincide at times shorter than $\tp=\td\Dt/\Da=\tau_0(\sig/\eps)^2$, referred to as the passive regime. This shows that the effect of the active fluctuations is hidden as long as the typical distance covered by the cage $\da\sqrt{t/\tau_0}$ is smaller than the cage size $\sig$. Between the two diffusions, a plateau regime appears when $\Da\ll\Dt$, as an evidence of the cage effect, and we observe a transient superdiffusion if $\Da\gg\Dt$. The time when the MSD deviates from the plateau, equal to $\tp$, can be shifted to an arbitrary large value. Conversely, the time when superdiffusion arises, also equal to $\tp$, can be arbitrarily short~\cite{Rao:14}. Our model contains the existence of ballistic directed events, as assessed by the superdiffusion, even if there is no persistence time in its formulation. The asymptotic behaviors of the MSD are summarized in Table~\ref{tab:sum}.

Beyond MSD, the 4-th moment is generally investigated to identify non-Gaussian features of the displacement statistics. More precisely, the deviation of the displacement distribution from Gaussian is quantified by the non-Gaussian parameter:
 \begin{equation}
\text{NGP}=\frac{\avg{\Delta x^4}}{3 \avg{\Delta x^2}^2} -1 \,\,.
\end{equation}
In our model, deviations from Gaussian behavior are governed by atypical events in which rare but important excursions occur. In that respect, the NGP characterizes the amount of directed events in the particle trajectory, probing the structural rearrangements of the system. Again using (\ref{eq:moments}) its calculation is straighforward:
\begin{equation}
\text{NGP} = \f{\tau_0}{3\td} \f{ 2\ee^{-\f{3t}{\td}} - 9 \ee^{-\f{2t}{\td}} + 18\ee^{-\f{t}{\td}} -11 + \f{6t}{\td} }{\brt{\pnt{\Dt/\Da-1} \ft + t/\td}^2}.
\end{equation}
Its evolution is determined by $\{\Da/\Dt,\td,\tau_0\}$, where $\tau_0$ only affects the amplitude. The NGP vanishes at short and long time, corresponding to the passive and active Gaussian regimes, respectively. It takes positive values in the transient regime, for which the distribution is broader than Gaussian [Fig.~\ref{fig:panels}(b)]. When $\Da\ll\Dt$, the peak time equals $t^*$, namely the time when the MSD deviates from the transient plateau, as observed in colloidal systems~\cite{Weeks:00,Weeks:02,Weeks:07,Weeks:15,Kilfoil:09}, and the peak value reads $(\eps/\sig)^2/2$. When $\Da\gg\Dt$, there is a plateau regime around the peak value, equal to $2\tau_0/\td$. In both cases, the NGP starts to decrease when the long time diffusive regime sets in. These asymptotic behaviors are also summarized in Table~\ref{tab:sum}.

\begin{table}
\caption{\label{tab:sum}
Different regimes in the time evolution of the MSD and the NGP.}
\centering
\begin{ruledtabular}
\begin{tabular}{cccc}
$\Da\ll\Dt$ & $t\ll\td$ & $\td\ll t\ll\td\sqrt{\Dt/\Da}$ & $\td\Dt/\Da \ll t$
\\
\colrule
\\
MSD & $2\Dt t$ & $2\Dt\td$ & $2\Da t$
\\\\
NGP & $\df{\tau_0}{2\td} \pnt{\df{\Da t}{\Dt \td} }^2$ & $2 t \tau_0 \pnt{\df{\Da}{\Dt \td} }^2$ & $\df{2\tau_0}{t}$
\\\\
\hline
\hline
$\Da\gg\Dt$ & $t\ll\td\Dt/\Da$ & $\td\Dt/\Da\ll t\ll\td$ & $\td\ll t$
\\
\colrule
\\
MSD & $2\Dt t$ & $\df{2\Da t^2}{\td}$ & $2\Da t$
\\\\
NGP & $\df{\tau_0}{2\td} \pnt{\df{\Da t}{\Dt \td} }^2$ & $\df{2\tau_0}{\td}$ & $\df{2\tau_0}{t}$
\\\\
\end{tabular}
\end{ruledtabular}
\end{table}


\subsection{Distribution of displacement}

To characterize further the non-Gaussian behavior of the displacement fluctuations, we have to deal with the complete form of the spatial Fourier transform (\ref{eq:tP}). Even though a complete Fourier Inversion is not possible analitycally, we can deduce several useful informations by looking at limiting cases, where approximations make calculations easier.

First, we observe that the active part of distribution in Eq.~(\ref{eq:tP}) can be simplified in the small and large $q$ limits
\begin{eqnarray}
\label{eq:Paas}
\tilde P_\text{\tiny A}(q,t) &\underset{q\da\ll1}{\sim}& \exp\brt{-(q\da)^2 \f{\td}{\tau_0} \pnt{ \f{t}{\td} - \ft }},
\\
\label{eq:Paag}
\tilde P_\text{\tiny A}(q,t) &\underset{q\da\gg1}{\sim}& \ee^{-t/\tau_0} \pnt{\abs{q}\da\ft}^{-\td/\tau_0}.
\end{eqnarray}
Taking $q=1/\sigma$, it follows that the total distribution $\tilde P$ is Gaussian at all times when $\da\ll\sigma$, whereas the non-Gaussian features affect the large time relaxation when $\da\gtrsim\sigma$, as reported in Tab.~\ref{tab:F}.

The displacement distribution is also always Gaussian at asymptotically short and long times, corresponding respectively to the passive and active diffusions. In the intermediate transient regime, it is Gaussian for small and large displacements, with a non-Gaussian crossover in between.

The active distribution $P_\text{\tiny A}$ has a Gaussian behavior for large displacements and large times, as a signature of the central limit theorem. It behaves like a power-law for small displacements with an exponential cut-off of the form $\ee^{- \abs{ \Delta x } / ( \eps\ft )}$. Mathematically, the exponential tails are due from the pole in the last term of (\ref{eq:tP}). Therefore, the total distribution $\tilde P$ is Gaussian at short and large displacement with exponential tails in between. The power-law behavior is only observed in the deviation from the central Gaussian part, as discussed below. To determine the deviation from the large Gaussian part, we expand $\ln \tilde P$ for small $q$ as
\begin{eqnarray}
\ln \tilde P(q,t) &=& - \f{ \avg{\Delta x^2(t)} }{2} q^2 + \f{ \avg{\Delta x^4(t)} - 3 \avg{\Delta x^2(t)}^2 }{4} q^4
\nonumber
\\
& & + \cO(q^6) .
\end{eqnarray}
It follows that the transition between the large Gaussian part and non-Gaussian features appears at $\Delta x \sim \brt{ \pnt{ \avg{ \Delta x^4(t) } / \avg{ \Delta x^2(t) } - 3 \avg{ \Delta x^2(t) } } / 2 }^{1/2} $.

\begin{table}[b]
\caption{\label{tab:F}
Different regimes in the time evolution of the Fourier displacement distribution at $q=1/\sig$. It is always Gaussian at times shorter than the terminal relaxation time $t^* = \td D / \Da$, and some non-Gaussian features appear at later times if $\da\gtrsim\sigma$.
}
\centering
\begin{ruledtabular}
\begin{tabular}{ccc}
$\tilde P(1/\sigma,t)$ & $\da\ll\sigma$  & $\da\gtrsim\sigma$
\\
\colrule
\\
$t\ll\ta$ & $\ee^{-(q\sig)^2 \ft}$ & $\ee^{-(q\sig)^2 \ft}$
\\\\
$t\gg\ta$ & $\ee^{-q^2 \Da t}$ & $\ee^{-(q\sig)^2 \ft - t/\tau_0} \pnt{\abs{q}\da\ft }^{-\td/\tau_0} $
\\\\
\end{tabular}
\end{ruledtabular}
\end{table}

The accessible range of displacements for experimental and simulated systems does not always allow one to observe the large Gaussian part of the distribution (they can clearly be seen, though, in \cite{Kilfoil:09}). The existence of directed events is quantified through the deviation from the central Gaussian, that is \textit{via} the cross-over between the two Gaussian parts. Within our model, it is given by exponential tails, as commonly reported in glassy systems~\cite{Berthier:07}, of the form $\ee^{ - \abs{\Delta x} / \pnt{ \eps \ft } }$. To quantitatitatively test the predictions of our model with experiments and computer simulations, we have compared our results with existing data for a dense suspension of colloidal particles~\cite{Kilfoil:09} and a binary Lennard-Jones mixture~\cite{Berthier:07, Berthier:07b}. Our model perfectly fits these results, as shown in Figure~\ref{fig:panels}(c) and (d). For the colloidal system, we reproduce the time evolution of the distribution with parameters $\cur{\tau_0/\td, \da/\sigma} = \cur{25, 6}$ [Fig.~\ref{fig:panels}(c)]. In the Lennard-Jones mixture, the measurements are taken for four temperatures at the $\alpha$ relaxation time, corresponding to $t^*$ within our model (see below). The central Gaussian part barely varies whereas the tails decrease with temperature. We identify the temperature with the diffusion coefficient $D$ of the confined motion, and we adjust the corresponding exponential tails with $\da = \cur{0.2, 0.25, 0.29, 0.33}$ from left to right in Fig.~\ref{fig:panels}(d). Available data supporting our choice of an exponential cage hop distribution however extend over a single decade, and it is likely that other choices of distribution could fit the data. More detail on the analysis on the experimental and numerical data is provided in Appendix~\ref{app:dataanalysis}.


To further characterize the departure from the central Gaussian, we investigate the transition between the central Gaussian part and the tails by looking at the intermediate displacement part of the distribution $P_\text{int}$ which does not include the exponential tails. We focus on the case $\tau_0 > \td$, namely when the time needed for the particle to relax within the cage is shorter than two successive cage hops.
We start considering the Fourier distribution of displacement at large wavenumber. From Eqs.~\eqref{eq:Ptt} and~\eqref{eq:Paag}, we deduce the expression of this distribution in the limit $q\da\gg1$
\begin{equation}\label{eq:sP}
\tilde P(q,t) \underset{q\da\gg1}{\sim} \ee^{-(q\sig)^2\ft-t/\tau_0} \pnt{\abs{q}\da\ft}^{-\td/\tau_0}.
\end{equation}
We perform the inverse Fourier transform of Eq.~\eqref{eq:sP} to obtain 
\begin{equation}\label{eq:1F1}
P(x,t) \underset{x\ll\eps}{\sim} {_1}F_1\brt{\f{\tau_0-\td}{2\tau_0},\f{1}{2} ; -\f{(x/\sig)^2}{ 4 \ft } } \equiv P_\text{int}(x,t) ,
\end{equation}
where ${_1}F_1$ is the confluent hypergeometric function of the first kind.  It depends on the hopping statistics only \textit{via} the typical waiting time $\tau_0$ between two successive cage hops, thus being independent of the hopping length $\da$. Besides, this result remains unchanged for any hopping distribution. This suggests that the intermediate displacement distribution is an appropriate probe to reveal universal behavior in glassy systems.

\begin{figure}
\centering
\includegraphics[width=\columnwidth]{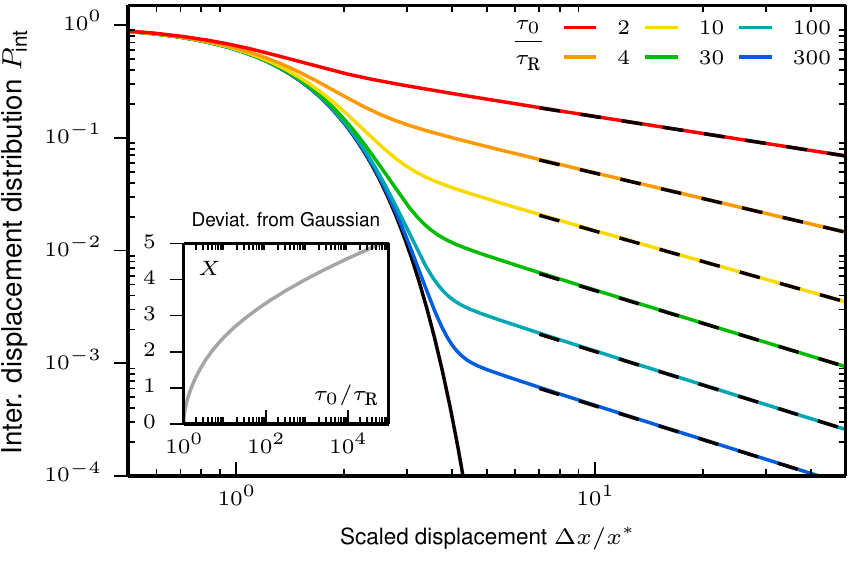}
\caption{\label{fig:sum}
Distribution of small displacement scaled by $x^*(t)=\sig\sqrt{2\ft/(1-\td/\tau_0)}$ in Eq.~\eqref{eq:1F1}. The central part is Gaussian (solid black line) with power-law tails (dashed lines). (Inset)~Cross-over value $X$ between the central Gaussian part and the tails of the distribution as a function of $\tau_0/\td$.
}
\end{figure}

The asymptotic behaviors of $P_\text{int}$ are given by
\begin{eqnarray}
\label{eq:1F1s}
P_\text{int} (x,t) &\underset{x\to0}{\sim}& \ee^{-\pnt{1-\f{\td}{\tau_0}} \f{(x/\sig)^2}{ 4 \ft }},
\\
\label{eq:1F1l}
P_\text{int} (x,t) &\underset{x\to\infty}{\sim}& \brt{\pnt{1-\f{\td}{\tau_0}} \f{(x/\sig)^2}{ 4 \ft } }^{\f{\td-\tau_0}{2\tau_0}} g_{\tau_0/\td},
\nonumber
\\
\end{eqnarray}
where
\begin{equation}
g_u =\f{1}{\sqrt{\pi}} \Gamma\pnt{1-\f{1}{2u}} \sin\pnt{\f{\pi}{2u}}.
\end{equation}
Therefore, the central part is Gaussian with standard deviation $\sig\sqrt{2\ft/(1-\td/\tau_0)}$, and the departure from this Gaussian is given by power-law tails with exponent $\td/\tau_0-1$ [Fig.~\ref{fig:sum}]. Note that $P_\text{int}$ only characterizes the deviation from the central Gaussian part, so that it does contain the exponential tails appearing for larger displacement. The typical length scale $X$ of the crossover between the Gaussian central part of this distribution and the power-law tails can be obtained by considering the small displacement distribution for displacements scaled by $x^*(t)=\sig\sqrt{2\ft/(1-\td/\tau_0)}$. From Eqs.~\eqref{eq:1F1s} and~\eqref{eq:1F1l}, we determine its expression as
\begin{equation}\label{eq:X}
X(u) = \sqrt{\f{1-u}{u}W \brt{ -2 \pnt{g_u}^{\f{2u}{u-1}} } },
\end{equation}
where $W$ is the principal branch of the Lambert W function defined by
\begin{equation}
z = W(z)\ee^{W(z)}.
\end{equation}
We plot $X$ as a function of $\tau_0/\td$ in the inset of Fig.~\ref{fig:sum}.


\section{Scattering function}

The comparison with glassy dynamics can be extended further by investigating the time evolution of the Fourier displacement distribution $\tilde P(q,t)$ at some fixed wavenumber, yet to be determined. To this end,we can regard $\tilde P(q,t)$ as an approximation for the sISF in a $N$-body system,
\begin{equation}
F_\text{\tiny S}(q,t) = \f{1}{N} \sum_{i=1}^N \avg{  \ee^{\ii q \brt{x_i(t) -x_i(0)}} }.
\end{equation}
In that case, the relevant choice for $q$ would be the value at the first peak of the structure factor, which corresponds to the inverse of the typical interparticle distance. Provided that caging stems from the steric hindrance, the typical distance between particles should be encoded in the cage size. This leads us to choose $1/\sig$ as the appropriate wavenumber.

\begin{figure}
\centering
\includegraphics[width=\columnwidth]{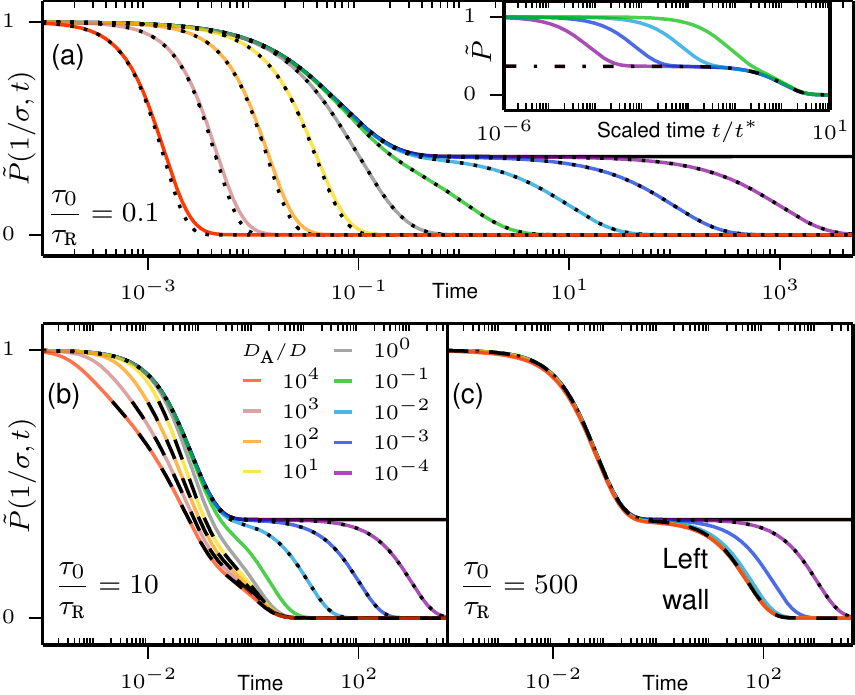}
\caption{\label{fig:F}
Flow curves of the Fourier displacement distribution at $q=1/\sigma$.
(a)~The transition from a single to a two-step relaxation as $\Da/\Dt$ decreases is the hallmark of glassy systems. There is no terminal relaxation for a steady cage (solid black line). The curves are Gaussian at all times when $\da\ll\sig$ (dotted lines).
(Inset)~For the two-step dynamics, we scale the time by the terminal relaxation time to reveal an exponential master curve (dot-dashed line). 
(b)~The non-Gaussian statistics affects the relaxation when $\da\gtrsim\sig$ (dashed lines).
(c)~The curves such as $\da\gtrsim\sig$ fall into the left wall $\ee^{- t/ \tau_0 + f_t}$ (dashed line), reflecting the dynamical slowing down as the cage hops become less frequent.
}
\end{figure}

There is a transition from a single to a two-step relaxation of the sISF as $\Da/\Dt$ decreases [Fig.~\ref{fig:F}]. This is typical of the behavior observed in glasses, the two-step dynamics being reminiscent of the $\beta$ and $\alpha$ relaxations. The sISF for a steady cage relaxes within a time $\td$ to a nonzero value, equal to the transient plateau value of the sISF for a hopping cage. Such behavior is similar to the kinetic arrest reported in glasses, when the particle evolves in a metastable state for an infinite long time. The plateau value reads $\ee^{-(q\sig)^2}$ for an arbitrary wavenumber. Cage rearrangements occur in the active case; then the particle overcomes the local confinement. The terminal relaxation for single and two-step dynamics starts at time $t^*$, namely when the passive regime ends. For a two-step dynamics, the terminal relaxation occurs when the NGP reaches its peak value~\cite{Kob:97,Kilfoil:09,Weeks:11,Weeks:15,Ikeda:12,Vogel}, and it can be shifted to an arbitrary long time~\cite{Stuart:13}.

The sISF is Gaussian at all times when $\eps\ll\sig$, the structural relaxation thus being entirely determined by the MSD. In such a case, the long time behavior is independent of $\Da$ when scaling the time by $\tp$. For a two-step relaxation, the corresponding master curve $\ee^{- (1+t/t^*)}$ is exponential [Inset of Fig.~\ref{fig:F}(a)]. A stretched exponential is usually reported, with an exponent close to one~\cite{Barrat:02,Berthier:02}. Our result suggests that one would clearly observe exponential behavior when the terminal relaxation time $t^*$ is large compared with the time $\td$ of the first relaxation. The non-Gaussian fluctuations play a role in the dynamics if $\da\gtrsim\sig$, the higher order statistics of the displacement thereby affecting the relaxation of the system [Fig.~\ref{fig:F}(b)].

A left wall appears in the flow curves if $\td\ll\tau_0$, namely all the flow curves such that $\da\gtrsim\sig$ fall onto a master curve [Fig.~\ref{fig:F}(c)]. The expression of the left wall reads $\ee^{-t/\tau_0 - \ft}$, the associated terminal relaxation time being $\tau_0$. The existence of a left wall is reminiscent of the dynamics for \textit{in silico} shear glasses~\cite{Berthier:03}. The sheared region close to the left wall reflects the slowing down of the relaxation as the cage hops become less frequent, that is when $\tau_0$ increases. On the opposite, the particle does not relax towards the center of the cage between two successive hops when $\td\gtrsim\tau_0$, in which case there is no left wall because the transient caging does not lead to any dynamical slowing down.


\section{Relation with many-body systems}

We now try to understand how our one-body approach can be connected with multicomponent systems. 
Using the expression of the Fourier displacement distribution in Eq.~(\ref{eq:tP}), we can deduce its time evolution
\begin{equation}\label{eq:fq}
\p_t \tilde P(q,t) = -q^2 \brt{ \Dt \ee^{-t/\td} + \f{\Da\ft}{1+\pnt{q\da\ft}^2 } } \tilde P(q,t).
\end{equation}
It follows that the displacement distribution obeys the following master equation
\begin{eqnarray}\label{eq:f}
\p_t P(x,t) &=&  \f{\rho_\text{ref}}{\zeta} \p_{xx}^2 \int \dd y V_\text{eff}(x-y,t) P(y,t) 
\nonumber
\\
& &+ D_\text{eff} (t) \p_{xx}^2 P(x,t),
\end{eqnarray}
where for later convenience we have introduced a reference density $\rho_\text{ref}$ and a friction coefficient $\zeta$. The time-dependent diffusion coefficient $D_\text{eff}$ and effective potential $V_\text{eff}$ read
\begin{eqnarray}
\label{eq:Deff}
D_\text{eff}(t) &=& \Dt \ee^{-t/\td},
\\
\label{eq:Veff}
V_\text{eff}(x,t) &=& \f{\zeta\da}{2 \rho_\text{ref} \tau_0} \exp \pnt{-\f{\abs{x}}{\da\ft}} ,
\end{eqnarray}
Our aim is to establish that the one-body problem can be mapped into a system of $N$ particles with diffusion coefficient $D_\text{eff}$ and interaction potential $V_\text{eff}$ which are time-dependent. The corresponding dynamics of a particle $i$ reads
\begin{equation}\label{eq:dyng}
\f{\dd x_i}{\dd t} = - \f{1}{\zeta} \sum_{j=1}^N \p_{x_i} V_\text{eff}(x_i-x_j(t),t) + \sqrt{2D_\text{eff}(t)}\xi_i(t),
\end{equation}
where $\xi_i$ is a zero-mean Gaussian noise with correlations $\avg{\xi_i(t)\xi_j(t')}=\delta_{ij}\delta(t-t')$. Note that Eq.~\eqref{eq:dyng} describes a nonequilibrium dynamics since the correlations of the noise are different from the prediction of the fluctuation-dissipation theorem. The self-intermediate scattering function $\Fs(q,t)$ is defined as
\begin{equation}
\Fs(q,t) = \f{1}{N} \sum_{i=1}^N \avg{  \ee^{\ii q \pnt{x_i(t) -x_i(0)}} }.
\end{equation}
The inverse Fourier transform of $\Fs(q,t)$ is the self-part of the Van Hove correlation function $\Gs(x,t)$
\begin{equation}
\Gs(x,t) = \f{1}{N}  \sum_{i=1}^N \avg{\delta(x+x_i(0)-x_i(t))}.
\end{equation}
The $n$-body correlation functions of the density are generally connected \textit{via} a hierarchy of equations. The simplest closure scheme, referred to as the random phase approximation (RPA), consists in truncating this hierarchy at the Gaussian order
\begin{eqnarray}\label{eq:VH}
\p_t \Gs(x,t) &=& \f{\rho_0}{\zeta} \p_{xx}^2 \int \dd y V_\text{eff}(x-y,t) \Gs(y,t)
\nonumber
\\
& &+ D_\text{eff}(t) \p_{xx}^2 \Gs(x,t),
\end{eqnarray}
where $\rho_0=N/L$ is defined in terms of the system size $L$. We identify the density of particle $\rho_0$ with the reference density $\rho_\text{ref}$ introduced earlier, and we choose $D_\text{eff}$ and $V_\text{eff}$ to obey Eqs.~\eqref{eq:Deff} and~\eqref{eq:Veff}, respectively. It follows that Eqs.~\eqref{eq:VH} and~\eqref{eq:f} describe the same dynamics, from which we deduce that the distribution of displacement for the one-body dynamics equals the Van Hove correlation function for interacting particles within the RPA. Likely, the Fourier displacement distribution corresponds to the self-intermediate scattering function. The effective potential arising from the non-Gaussian statistics of the cage's dynamics can be regarded as a mean-field potential resulting from the interaction between many particles. Such potential describes repulsive interactions with a very soft core, thus allowing the overlap of the particles with a finite cost of energy~\cite{Dyre}. The range of interaction increases with time as $\ft$, namely it saturates within a time $\td$ to $\da$. Therefore, there is a transition between: (i) a diffusive dynamics with very short interaction, (ii) a dynamics with a very small diffusion coefficient and a larger interaction range, corresponding respectively to times shorter and longer than $\td$. The existence of interactions is a direct consequence of the cage rearrangements within our model. The explicit form of the potential is determined by the statistics of the cage hops, which are exponentially distributed in the present case. This suggests that the non-Gaussian properties of the particle dynamics hold information about the details of interaction between the particles.


\section{Discussion}

We have presented a minimal model of a particle immersed in a glassy system, based on an active cage with non-Gaussian dynamics. By focussing on a single particle we are able to decouple complex phenomena that arise from collective effects from the sole dynamics of the particle. Of course this approach is not intended in understanding how collective phenomena emerge, but rather to provide a vivid picture explaining experiments on glassy systems. Despite the simplicity of this model, we have demonstrated that its dynamics encompasses the complex behaviors arising in glassy systems, and that the onset of glassy behavior is shifted by the active component of the dynamics, in line with numerical evidence~\cite{Stuart:13,Berthier:14,Rao:14,Gomper:14}. Moreover, we have highlighted the scale invariance of the small displacement distribution. The dynamics of tracers in living systems shares common features with the glassy dynamics~\cite{Bursac,Hannezo:11}. The tracers can be either attached or embedded in a network of filaments, whose rearrangement is induced by nonequilibrium processes, thus monitoring transitions between locally stable configurations~\cite{Weitz:09, Ahmed:15}. The minimal approach of intermittent dynamics that we offer in the present paper should also be relevant for a large variety of biological systems~\cite{Mizuno:14a, Mizuno:14b, Kanazawa:14, Visco:15, Gov:15, Riveline:15, Ahmed:15b, Ahmed:15c}, as a useful tool to understand the existence of a scale-invariant regime in the tracer displacement~\cite{Bursac,Weihs,Granick}.

\begin{acknowledgments}
The Authors acknowledge useful discussions with Kiyoshi Kanazawa and Daisuke Mizuno.
\'EF and FvW acknowledge the support of the YITP and SoftActive program, as well as of the JSPS core-to-core program ``Non-equilibrium dynamics of soft matter and information". This work was partially supported by JSPS KAKENHI Grant No. 25287098 and 16H04025.
\end{acknowledgments}
\appendix

\section{Generalization to an arbitrary hopping distribution}
\label{app:nongaussian}
We introduce the distribution $\cW$ which prescribes the distance of the instantaneous hops experienced by the cage. We assume that the waiting time between time two consecutive jumps remains exponentially distributed with mean value $\tau_0$. Hence, the cumulants of $\va$ read
\begin{equation}\label{eq:vab}
\avgc{\va(t_1)\dots \va(t_n)} = K_n\delta_n(t_1,\dots, t_n),
\end{equation}
where
\begin{equation}\label{eq:Kn}
K_n=\f{1}{\tau_0}\int \dd x_0 \cW(x_0) x_0^n.
\end{equation}
Note that we recover Eq.~\eqref{eq:va} for an exponential distribution $\cW(x_0) = \ee^{-\abs{x_0}/\da}/\da$, as expected. We express the active distribution in terms of the cumulant coefficients $K_n$ as
\begin{equation}\label{eq:Paab}
\tilde P_\text{\tiny A}(q,t) =\exp\brt{\sum_{n=1}^\infty\f{(\ii q)^n K_n}{n!} \pnt{ \f{\ft^n}{n} + \int_0^t \dd s f_{t-s}^n } }.
\end{equation}
We deduce the time-derivative of the total distribution
\begin{eqnarray}\label{eq:Pq}
\p_t \tilde P(q,t) &=&  -q^2 D_\text{eff}(t) \tilde P(q,t)
\nonumber
\\
& &+ \sum_{n=1}^\infty\f{(\ii q)^n K_n}{n!} \f{\dd}{\dd t} \pnt{\f{\ft^n}{n} + \int_0^t \dd s f_{t-s}^n } \tilde P(q,t) ,
\nonumber
\\
\end{eqnarray}
where
\begin{equation}\label{eq:f}
\f{\dd}{\dd t} \pnt{ \f{\ft^n}{n} + \int_0^t \dd s f_{t-s}^n } = \ee^{-t/\td} \ft^{n-1} + \ft^n = \ft^{n-1}.
\end{equation}
By using Eqs.~\eqref{eq:Kn} and~\eqref{eq:f}, we write the second term in the right-hand side of Eq.~\eqref{eq:Pq} in terms of the hopping distribution $\cW$ as
\begin{eqnarray}\label{eq:fqb}
\brt{ \p_t + q^2 D_\text{eff} } \tilde P &=&  \f{1}{\tau_0 \ft} \int \dd x_0 \cW(x_0)\sum_{n=1}^\infty\f{(\ii q x_0)^n}{n!} \ft^n \tilde P
\nonumber
\\
\\
&=&\f{1}{\tau_0 \ft} \int \dd x_0 \cW(x_0) \brt{ \ee^{\ii q x_0 \ft}-1 } \tilde P
\nonumber
\\
\\
\label{eq:fqc}
&=&\f{1}{\tau_0 \ft} \brt{ \tilde \cW\pnt{q\ft} - 1 } \tilde P,
\end{eqnarray}
where we have used $\int \dd x_0 \cW(x_0)=1$. Our aim lies in deriving the distribution for small displacements. To this end, we consider Eq.~\eqref{eq:fqc} in the large $q$ limit. Provided that the hopping distribution $\cW$ is not a delta function, its Fourier transform should decay to zero for wavenumbers sufficiently large compared with the inverse typical hopping length $1/\da$. In this limit, we express Eq.~\eqref{eq:fqc} as
\begin{equation}
\p_t \tilde P(q,t) \underset{q\da\gg1}{\sim} - \brt{ q^2 D_\text{eff}(t) + \f{1}{\tau_0 \ft} } \tilde P(q,t).
\end{equation}
The solution of the above equation is consistent with Eq.~\eqref{eq:sP}. However, the power-law prefactor in $q$ in Eq.~\eqref{eq:sP} can only be obtained after a careful analysis of the asymptotics of $\tilde{\mathcal{W}}(q f_t)$ (and it shows up only if $\mathcal{W}$ decays exponentially or slower at large distances).


\section{Data analysis}
\label{app:dataanalysis}
We present in this appendix the analysis of the distribution of displacement obtained from two different systems: (i) a binary Lennard-Jones glass-forming mixture~\cite{Berthier:07, Berthier:07b}, and (ii) a dense suspension of colloid particles~\cite{Kilfoil:09}. In the two systems, we can reproduce both the central Gaussian part and the exponential tails as reported in Figs.~\ref{fig:panels}(c,d).

For the Lennard-Jones mixture, we scale the displacement distribution by the standard deviation of the Gaussian central part. The measurements are taken for four temperatures at the $\alpha$ relaxation time, corresponding to $t^* = \td D/\Da$ within our model. We identify the temperature with the diffusion coefficient $D$ of the confined motion, letting us with three free parameters $\cur{\td, \tau_0, \da}$. The time scales $\cur{\td, \tau_0} = \cur{0.05, 1.05}$ are taken the same for all temperatures, and we adjust the corresponding exponential tails with $\da$.

For the colloidal system, the measurements are taken at five different times. We fix the ratio $\td/\tau_0$ from the deviation of the central Gaussian part, and we determine $\cur{ \sigma, \td, \tau_0 } = \cur{ 0.05\,\mu\text{m}, 10^3\,\text{s}, 25\times10^3\,\text{s} }$ by fitting the central Gaussian parts of the distribution. Eventually, we adjust the exponential tails with $\da = 0.3$~$\mu$m. The time evolution of the distribution is reproduced with the same set of parameters.

\bibliographystyle{apsrev4-1}
\bibliography{draft-ref}

\begin{thebibliography}{62}%
\makeatletter
\providecommand \@ifxundefined [1]{%
 \@ifx{#1\undefined}
}%
\providecommand \@ifnum [1]{%
 \ifnum #1\expandafter \@firstoftwo
 \else \expandafter \@secondoftwo
 \fi
}%
\providecommand \@ifx [1]{%
 \ifx #1\expandafter \@firstoftwo
 \else \expandafter \@secondoftwo
 \fi
}%
\providecommand \natexlab [1]{#1}%
\providecommand \enquote  [1]{``#1''}%
\providecommand \bibnamefont  [1]{#1}%
\providecommand \bibfnamefont [1]{#1}%
\providecommand \citenamefont [1]{#1}%
\providecommand \href@noop [0]{\@secondoftwo}%
\providecommand \href [0]{\begingroup \@sanitize@url \@href}%
\providecommand \@href[1]{\@@startlink{#1}\@@href}%
\providecommand \@@href[1]{\endgroup#1\@@endlink}%
\providecommand \@sanitize@url [0]{\catcode `\\12\catcode `\$12\catcode
  `\&12\catcode `\#12\catcode `\^12\catcode `\_12\catcode `\%12\relax}%
\providecommand \@@startlink[1]{}%
\providecommand \@@endlink[0]{}%
\providecommand \url  [0]{\begingroup\@sanitize@url \@url }%
\providecommand \@url [1]{\endgroup\@href {#1}{\urlprefix }}%
\providecommand \urlprefix  [0]{URL }%
\providecommand \Eprint [0]{\href }%
\providecommand \doibase [0]{http://dx.doi.org/}%
\providecommand \selectlanguage [0]{\@gobble}%
\providecommand \bibinfo  [0]{\@secondoftwo}%
\providecommand \bibfield  [0]{\@secondoftwo}%
\providecommand \translation [1]{[#1]}%
\providecommand \BibitemOpen [0]{}%
\providecommand \bibitemStop [0]{}%
\providecommand \bibitemNoStop [0]{.\EOS\space}%
\providecommand \EOS [0]{\spacefactor3000\relax}%
\providecommand \BibitemShut  [1]{\csname bibitem#1\endcsname}%
\let\auto@bib@innerbib\@empty
\bibitem [{\citenamefont {Berthier}\ and\ \citenamefont
  {Biroli}(2011)}]{Biroli:11}%
  \BibitemOpen
  \bibfield  {author} {\bibinfo {author} {\bibfnamefont {L.}~\bibnamefont
  {Berthier}}\ and\ \bibinfo {author} {\bibfnamefont {G.}~\bibnamefont
  {Biroli}},\ }\href {\doibase 10.1103/RevModPhys.83.587} {\bibfield  {journal}
  {\bibinfo  {journal} {Rev. Mod. Phys.}\ }\textbf {\bibinfo {volume} {83}},\
  \bibinfo {pages} {587} (\bibinfo {year} {2011})}\BibitemShut {NoStop}%
\bibitem [{\citenamefont {G\"otze}(2009)}]{Gotze}%
  \BibitemOpen
  \bibfield  {author} {\bibinfo {author} {\bibfnamefont {W.}~\bibnamefont
  {G\"otze}},\ }\href@noop {} {\emph {\bibinfo {title} {Complex dynamics of
  glass-forming liquids, a mode-coupling theory}}}\ (\bibinfo  {publisher}
  {Oxford University Press},\ \bibinfo {address} {Oxford},\ \bibinfo {year}
  {2009})\BibitemShut {NoStop}%
\bibitem [{\citenamefont {Janssen}\ and\ \citenamefont
  {Reichman}(2015)}]{PhysRevLett.115.205701}%
  \BibitemOpen
  \bibfield  {author} {\bibinfo {author} {\bibfnamefont {L.~M.~C.}\
  \bibnamefont {Janssen}}\ and\ \bibinfo {author} {\bibfnamefont {D.~R.}\
  \bibnamefont {Reichman}},\ }\href {\doibase 10.1103/PhysRevLett.115.205701}
  {\bibfield  {journal} {\bibinfo  {journal} {Phys. Rev. Lett.}\ }\textbf
  {\bibinfo {volume} {115}},\ \bibinfo {pages} {205701} (\bibinfo {year}
  {2015})}\BibitemShut {NoStop}%
\bibitem [{\citenamefont {{Priya}}\ \emph {et~al.}(2015)\citenamefont
  {{Priya}}, \citenamefont {{Bidhoodi}},\ and\ \citenamefont
  {{Das}}}]{2015arXiv151100254P}%
  \BibitemOpen
  \bibfield  {author} {\bibinfo {author} {\bibfnamefont {M.}~\bibnamefont
  {{Priya}}}, \bibinfo {author} {\bibfnamefont {N.}~\bibnamefont {{Bidhoodi}}},
  \ and\ \bibinfo {author} {\bibfnamefont {S.~P.}\ \bibnamefont {{Das}}},\
  }\href@noop {} {\bibfield  {journal} {\bibinfo  {journal} {ArXiv e-prints}\ }
  (\bibinfo {year} {2015})}\BibitemShut {NoStop}%
\bibitem [{\citenamefont {{Bidhoodi}}\ and\ \citenamefont
  {{Das}}(2015)}]{2015arXiv151100786B}%
  \BibitemOpen
  \bibfield  {author} {\bibinfo {author} {\bibfnamefont {N.}~\bibnamefont
  {{Bidhoodi}}}\ and\ \bibinfo {author} {\bibfnamefont {S.~P.}\ \bibnamefont
  {{Das}}},\ }\href@noop {} {\bibfield  {journal} {\bibinfo  {journal} {ArXiv
  e-prints}\ } (\bibinfo {year} {2015})}\BibitemShut {NoStop}%
\bibitem [{\citenamefont {Weeks}\ \emph {et~al.}(2000)\citenamefont {Weeks},
  \citenamefont {Crocker}, \citenamefont {Levitt}, \citenamefont {Schofield},\
  and\ \citenamefont {Weitz}}]{Weeks:00}%
  \BibitemOpen
  \bibfield  {author} {\bibinfo {author} {\bibfnamefont {E.~R.}\ \bibnamefont
  {Weeks}}, \bibinfo {author} {\bibfnamefont {J.~C.}\ \bibnamefont {Crocker}},
  \bibinfo {author} {\bibfnamefont {A.~C.}\ \bibnamefont {Levitt}}, \bibinfo
  {author} {\bibfnamefont {A.}~\bibnamefont {Schofield}}, \ and\ \bibinfo
  {author} {\bibfnamefont {D.~A.}\ \bibnamefont {Weitz}},\ }\href {\doibase
  10.1126/science.287.5453.627} {\bibfield  {journal} {\bibinfo  {journal}
  {Science}\ }\textbf {\bibinfo {volume} {287}},\ \bibinfo {pages} {627}
  (\bibinfo {year} {2000})}\BibitemShut {NoStop}%
\bibitem [{\citenamefont {Kegel}\ and\ \citenamefont {van
  Blaaderen}(2000)}]{Kegel:00}%
  \BibitemOpen
  \bibfield  {author} {\bibinfo {author} {\bibfnamefont {W.~K.}\ \bibnamefont
  {Kegel}}\ and\ \bibinfo {author} {\bibfnamefont {A.}~\bibnamefont {van
  Blaaderen}},\ }\href {\doibase 10.1126/science.287.5451.290} {\bibfield
  {journal} {\bibinfo  {journal} {Science}\ }\textbf {\bibinfo {volume}
  {287}},\ \bibinfo {pages} {290} (\bibinfo {year} {2000})}\BibitemShut
  {NoStop}%
\bibitem [{\citenamefont {Weeks}\ and\ \citenamefont {Weitz}(2002)}]{Weeks:02}%
  \BibitemOpen
  \bibfield  {author} {\bibinfo {author} {\bibfnamefont {E.~R.}\ \bibnamefont
  {Weeks}}\ and\ \bibinfo {author} {\bibfnamefont {D.~A.}\ \bibnamefont
  {Weitz}},\ }\href {\doibase 10.1103/PhysRevLett.89.095704} {\bibfield
  {journal} {\bibinfo  {journal} {Phys. Rev. Lett.}\ }\textbf {\bibinfo
  {volume} {89}},\ \bibinfo {pages} {095704} (\bibinfo {year}
  {2002})}\BibitemShut {NoStop}%
\bibitem [{\citenamefont {Gao}\ and\ \citenamefont
  {Kilfoil}(2007)}]{Kilfoil:07}%
  \BibitemOpen
  \bibfield  {author} {\bibinfo {author} {\bibfnamefont {Y.}~\bibnamefont
  {Gao}}\ and\ \bibinfo {author} {\bibfnamefont {M.~L.}\ \bibnamefont
  {Kilfoil}},\ }\href {\doibase 10.1103/PhysRevLett.99.078301} {\bibfield
  {journal} {\bibinfo  {journal} {Phys. Rev. Lett.}\ }\textbf {\bibinfo
  {volume} {99}},\ \bibinfo {pages} {078301} (\bibinfo {year}
  {2007})}\BibitemShut {NoStop}%
\bibitem [{\citenamefont {Gao}\ and\ \citenamefont
  {Kilfoil}(2009{\natexlab{a}})}]{Kilfoil:09}%
  \BibitemOpen
  \bibfield  {author} {\bibinfo {author} {\bibfnamefont {Y.}~\bibnamefont
  {Gao}}\ and\ \bibinfo {author} {\bibfnamefont {M.~L.}\ \bibnamefont
  {Kilfoil}},\ }\href {\doibase 10.1103/PhysRevE.79.051406} {\bibfield
  {journal} {\bibinfo  {journal} {Phys. Rev. E}\ }\textbf {\bibinfo {volume}
  {79}},\ \bibinfo {pages} {051406} (\bibinfo {year}
  {2009}{\natexlab{a}})}\BibitemShut {NoStop}%
\bibitem [{\citenamefont {Caswell}\ \emph {et~al.}(2013)\citenamefont
  {Caswell}, \citenamefont {Zhang}, \citenamefont {Gardel},\ and\ \citenamefont
  {Nagel}}]{Gardel:13}%
  \BibitemOpen
  \bibfield  {author} {\bibinfo {author} {\bibfnamefont {T.~A.}\ \bibnamefont
  {Caswell}}, \bibinfo {author} {\bibfnamefont {Z.}~\bibnamefont {Zhang}},
  \bibinfo {author} {\bibfnamefont {M.~L.}\ \bibnamefont {Gardel}}, \ and\
  \bibinfo {author} {\bibfnamefont {S.~R.}\ \bibnamefont {Nagel}},\ }\href
  {\doibase 10.1103/PhysRevE.87.012303} {\bibfield  {journal} {\bibinfo
  {journal} {Phys. Rev. E}\ }\textbf {\bibinfo {volume} {87}},\ \bibinfo
  {pages} {012303} (\bibinfo {year} {2013})}\BibitemShut {NoStop}%
\bibitem [{\citenamefont {Stariolo}\ and\ \citenamefont
  {Fabricius}(2006)}]{Stariolo:06}%
  \BibitemOpen
  \bibfield  {author} {\bibinfo {author} {\bibfnamefont {D.~A.}\ \bibnamefont
  {Stariolo}}\ and\ \bibinfo {author} {\bibfnamefont {G.}~\bibnamefont
  {Fabricius}},\ }\href {\doibase 10.1063/1.2221309} {\bibfield  {journal}
  {\bibinfo  {journal} {J. Chem. Phys.}\ }\textbf {\bibinfo {volume} {125}},\
  \bibinfo {pages} {064505} (\bibinfo {year} {2006})}\BibitemShut {NoStop}%
\bibitem [{\citenamefont {Chaudhuri}\ \emph {et~al.}(2007)\citenamefont
  {Chaudhuri}, \citenamefont {Berthier},\ and\ \citenamefont
  {Kob}}]{Berthier:07}%
  \BibitemOpen
  \bibfield  {author} {\bibinfo {author} {\bibfnamefont {P.}~\bibnamefont
  {Chaudhuri}}, \bibinfo {author} {\bibfnamefont {L.}~\bibnamefont {Berthier}},
  \ and\ \bibinfo {author} {\bibfnamefont {W.}~\bibnamefont {Kob}},\ }\href
  {\doibase 10.1103/PhysRevLett.99.060604} {\bibfield  {journal} {\bibinfo
  {journal} {Phys. Rev. Lett.}\ }\textbf {\bibinfo {volume} {99}},\ \bibinfo
  {pages} {060604} (\bibinfo {year} {2007})}\BibitemShut {NoStop}%
\bibitem [{\citenamefont {Gao}\ and\ \citenamefont
  {Kilfoil}(2009{\natexlab{b}})}]{Gao:09}%
  \BibitemOpen
  \bibfield  {author} {\bibinfo {author} {\bibfnamefont {Y.}~\bibnamefont
  {Gao}}\ and\ \bibinfo {author} {\bibfnamefont {M.~L.}\ \bibnamefont
  {Kilfoil}},\ }\href {\doibase 10.1364/OE.17.004685} {\bibfield  {journal}
  {\bibinfo  {journal} {Opt. Express}\ }\textbf {\bibinfo {volume} {17}},\
  \bibinfo {pages} {4685} (\bibinfo {year} {2009}{\natexlab{b}})}\BibitemShut
  {NoStop}%
\bibitem [{\citenamefont {Helfferich}\ \emph
  {et~al.}(2014{\natexlab{a}})\citenamefont {Helfferich}, \citenamefont
  {Ziebert}, \citenamefont {Frey}, \citenamefont {Meyer}, \citenamefont
  {Farago}, \citenamefont {Blumen},\ and\ \citenamefont
  {Baschnagel}}]{Farago:14a}%
  \BibitemOpen
  \bibfield  {author} {\bibinfo {author} {\bibfnamefont {J.}~\bibnamefont
  {Helfferich}}, \bibinfo {author} {\bibfnamefont {F.}~\bibnamefont {Ziebert}},
  \bibinfo {author} {\bibfnamefont {S.}~\bibnamefont {Frey}}, \bibinfo {author}
  {\bibfnamefont {H.}~\bibnamefont {Meyer}}, \bibinfo {author} {\bibfnamefont
  {J.}~\bibnamefont {Farago}}, \bibinfo {author} {\bibfnamefont
  {A.}~\bibnamefont {Blumen}}, \ and\ \bibinfo {author} {\bibfnamefont
  {J.}~\bibnamefont {Baschnagel}},\ }\href {\doibase
  10.1103/PhysRevE.89.042603} {\bibfield  {journal} {\bibinfo  {journal} {Phys.
  Rev. E}\ }\textbf {\bibinfo {volume} {89}},\ \bibinfo {pages} {042603}
  (\bibinfo {year} {2014}{\natexlab{a}})}\BibitemShut {NoStop}%
\bibitem [{\citenamefont {Helfferich}\ \emph
  {et~al.}(2014{\natexlab{b}})\citenamefont {Helfferich}, \citenamefont
  {Ziebert}, \citenamefont {Frey}, \citenamefont {Meyer}, \citenamefont
  {Farago}, \citenamefont {Blumen},\ and\ \citenamefont
  {Baschnagel}}]{Farago:14b}%
  \BibitemOpen
  \bibfield  {author} {\bibinfo {author} {\bibfnamefont {J.}~\bibnamefont
  {Helfferich}}, \bibinfo {author} {\bibfnamefont {F.}~\bibnamefont {Ziebert}},
  \bibinfo {author} {\bibfnamefont {S.}~\bibnamefont {Frey}}, \bibinfo {author}
  {\bibfnamefont {H.}~\bibnamefont {Meyer}}, \bibinfo {author} {\bibfnamefont
  {J.}~\bibnamefont {Farago}}, \bibinfo {author} {\bibfnamefont
  {A.}~\bibnamefont {Blumen}}, \ and\ \bibinfo {author} {\bibfnamefont
  {J.}~\bibnamefont {Baschnagel}},\ }\href {\doibase
  10.1103/PhysRevE.89.042604} {\bibfield  {journal} {\bibinfo  {journal} {Phys.
  Rev. E}\ }\textbf {\bibinfo {volume} {89}},\ \bibinfo {pages} {042604}
  (\bibinfo {year} {2014}{\natexlab{b}})}\BibitemShut {NoStop}%
\bibitem [{\citenamefont {Hedges}\ \emph {et~al.}(2007)\citenamefont {Hedges},
  \citenamefont {Maibaum}, \citenamefont {Chandler},\ and\ \citenamefont
  {Garrahan}}]{Chandler:07}%
  \BibitemOpen
  \bibfield  {author} {\bibinfo {author} {\bibfnamefont {L.~O.}\ \bibnamefont
  {Hedges}}, \bibinfo {author} {\bibfnamefont {L.}~\bibnamefont {Maibaum}},
  \bibinfo {author} {\bibfnamefont {D.}~\bibnamefont {Chandler}}, \ and\
  \bibinfo {author} {\bibfnamefont {J.~P.}\ \bibnamefont {Garrahan}},\ }\href
  {\doibase 10.1063/1.2803062} {\bibfield  {journal} {\bibinfo  {journal} {J.
  Chem. Phys.}\ }\textbf {\bibinfo {volume} {127}},\ \bibinfo {pages} {211101}
  (\bibinfo {year} {2007})}\BibitemShut {NoStop}%
\bibitem [{\citenamefont {Pastore}\ \emph {et~al.}(2014)\citenamefont
  {Pastore}, \citenamefont {Coniglio},\ and\ \citenamefont
  {Pica~Ciamarra}}]{Pastore:14}%
  \BibitemOpen
  \bibfield  {author} {\bibinfo {author} {\bibfnamefont {R.}~\bibnamefont
  {Pastore}}, \bibinfo {author} {\bibfnamefont {A.}~\bibnamefont {Coniglio}}, \
  and\ \bibinfo {author} {\bibfnamefont {M.}~\bibnamefont {Pica~Ciamarra}},\
  }\href {\doibase 10.1039/C4SM00739E} {\bibfield  {journal} {\bibinfo
  {journal} {Soft Matter}\ }\textbf {\bibinfo {volume} {10}},\ \bibinfo {pages}
  {5724} (\bibinfo {year} {2014})}\BibitemShut {NoStop}%
\bibitem [{\citenamefont {{Pastore}}\ \emph {et~al.}(2014)\citenamefont
  {{Pastore}}, \citenamefont {{Coniglio}},\ and\ \citenamefont {{Pica
  Ciamarra}}}]{Pastore:14b}%
  \BibitemOpen
  \bibfield  {author} {\bibinfo {author} {\bibfnamefont {R.}~\bibnamefont
  {{Pastore}}}, \bibinfo {author} {\bibfnamefont {A.}~\bibnamefont
  {{Coniglio}}}, \ and\ \bibinfo {author} {\bibfnamefont {M.}~\bibnamefont
  {{Pica Ciamarra}}},\ }\href {\doibase 10.1038/srep11770} {\bibfield
  {journal} {\bibinfo  {journal} {Sci. Rep.}\ }\textbf {\bibinfo {volume}
  {5}},\ \bibinfo {pages} {11770} (\bibinfo {year} {2014})}\BibitemShut
  {NoStop}%
\bibitem [{\citenamefont {Pastore}\ \emph {et~al.}(2015)\citenamefont
  {Pastore}, \citenamefont {Pica~Ciamarra}, \citenamefont {Pesce},\ and\
  \citenamefont {Sasso}}]{Pastore:15}%
  \BibitemOpen
  \bibfield  {author} {\bibinfo {author} {\bibfnamefont {R.}~\bibnamefont
  {Pastore}}, \bibinfo {author} {\bibfnamefont {M.}~\bibnamefont
  {Pica~Ciamarra}}, \bibinfo {author} {\bibfnamefont {G.}~\bibnamefont
  {Pesce}}, \ and\ \bibinfo {author} {\bibfnamefont {A.}~\bibnamefont
  {Sasso}},\ }\href {\doibase 10.1039/C4SM02147A} {\bibfield  {journal}
  {\bibinfo  {journal} {Soft Matter}\ }\textbf {\bibinfo {volume} {11}},\
  \bibinfo {pages} {622} (\bibinfo {year} {2015})}\BibitemShut {NoStop}%
\bibitem [{\citenamefont {Helfferich}(2014)}]{Helfferich}%
  \BibitemOpen
  \bibfield  {author} {\bibinfo {author} {\bibfnamefont {J.}~\bibnamefont
  {Helfferich}},\ }\href {\doibase 10.1140/epje/i2014-14073-6} {\bibfield
  {journal} {\bibinfo  {journal} {Eur. Phys. J. E}\ }\textbf {\bibinfo {volume}
  {37}} (\bibinfo {year} {2014}),\ 10.1140/epje/i2014-14073-6}\BibitemShut
  {NoStop}%
\bibitem [{\citenamefont {Chaudhuri}\ \emph {et~al.}(2008)\citenamefont
  {Chaudhuri}, \citenamefont {Gao}, \citenamefont {Berthier}, \citenamefont
  {Kilfoil},\ and\ \citenamefont {Kob}}]{Berthier:08}%
  \BibitemOpen
  \bibfield  {author} {\bibinfo {author} {\bibfnamefont {P.}~\bibnamefont
  {Chaudhuri}}, \bibinfo {author} {\bibfnamefont {Y.}~\bibnamefont {Gao}},
  \bibinfo {author} {\bibfnamefont {L.}~\bibnamefont {Berthier}}, \bibinfo
  {author} {\bibfnamefont {M.}~\bibnamefont {Kilfoil}}, \ and\ \bibinfo
  {author} {\bibfnamefont {W.}~\bibnamefont {Kob}},\ }\href {\doibase
  10.1088/0953-8984/20/24/244126} {\bibfield  {journal} {\bibinfo  {journal}
  {J. Phys.: Condens. Matter}\ }\textbf {\bibinfo {volume} {20}},\ \bibinfo
  {pages} {244126} (\bibinfo {year} {2008})}\BibitemShut {NoStop}%
\bibitem [{\citenamefont {Berthier}\ \emph {et~al.}(2000)\citenamefont
  {Berthier}, \citenamefont {Barrat},\ and\ \citenamefont
  {Kurchan}}]{Berthier:00}%
  \BibitemOpen
  \bibfield  {author} {\bibinfo {author} {\bibfnamefont {L.}~\bibnamefont
  {Berthier}}, \bibinfo {author} {\bibfnamefont {J.-L.}\ \bibnamefont
  {Barrat}}, \ and\ \bibinfo {author} {\bibfnamefont {J.}~\bibnamefont
  {Kurchan}},\ }\href {\doibase 10.1103/PhysRevE.61.5464} {\bibfield  {journal}
  {\bibinfo  {journal} {Phys. Rev. E}\ }\textbf {\bibinfo {volume} {61}},\
  \bibinfo {pages} {5464} (\bibinfo {year} {2000})}\BibitemShut {NoStop}%
\bibitem [{\citenamefont {Berthier}\ and\ \citenamefont
  {Barrat}(2002{\natexlab{a}})}]{Barrat:02}%
  \BibitemOpen
  \bibfield  {author} {\bibinfo {author} {\bibfnamefont {L.}~\bibnamefont
  {Berthier}}\ and\ \bibinfo {author} {\bibfnamefont {J.-L.}\ \bibnamefont
  {Barrat}},\ }\href {\doibase 10.1103/PhysRevLett.89.095702} {\bibfield
  {journal} {\bibinfo  {journal} {Phys. Rev. Lett.}\ }\textbf {\bibinfo
  {volume} {89}},\ \bibinfo {pages} {095702} (\bibinfo {year}
  {2002}{\natexlab{a}})}\BibitemShut {NoStop}%
\bibitem [{\citenamefont {Berthier}\ and\ \citenamefont
  {Barrat}(2002{\natexlab{b}})}]{Berthier:02}%
  \BibitemOpen
  \bibfield  {author} {\bibinfo {author} {\bibfnamefont {L.}~\bibnamefont
  {Berthier}}\ and\ \bibinfo {author} {\bibfnamefont {J.-L.}\ \bibnamefont
  {Barrat}},\ }\href {\doibase 10.1063/1.1460862} {\bibfield  {journal}
  {\bibinfo  {journal} {J. Chem. Phys.}\ }\textbf {\bibinfo {volume} {116}},\
  \bibinfo {pages} {6228} (\bibinfo {year} {2002}{\natexlab{b}})}\BibitemShut
  {NoStop}%
\bibitem [{\citenamefont {Varnik}\ \emph {et~al.}(2003)\citenamefont {Varnik},
  \citenamefont {Bocquet}, \citenamefont {Barrat},\ and\ \citenamefont
  {Berthier}}]{Berthier:03}%
  \BibitemOpen
  \bibfield  {author} {\bibinfo {author} {\bibfnamefont {F.}~\bibnamefont
  {Varnik}}, \bibinfo {author} {\bibfnamefont {L.}~\bibnamefont {Bocquet}},
  \bibinfo {author} {\bibfnamefont {J.-L.}\ \bibnamefont {Barrat}}, \ and\
  \bibinfo {author} {\bibfnamefont {L.}~\bibnamefont {Berthier}},\ }\href
  {\doibase 10.1103/PhysRevLett.90.095702} {\bibfield  {journal} {\bibinfo
  {journal} {Phys. Rev. Lett.}\ }\textbf {\bibinfo {volume} {90}},\ \bibinfo
  {pages} {095702} (\bibinfo {year} {2003})}\BibitemShut {NoStop}%
\bibitem [{\citenamefont {Ni}\ \emph {et~al.}(2013)\citenamefont {Ni},
  \citenamefont {Stuart},\ and\ \citenamefont {Dijkstra}}]{Stuart:13}%
  \BibitemOpen
  \bibfield  {author} {\bibinfo {author} {\bibfnamefont {R.}~\bibnamefont
  {Ni}}, \bibinfo {author} {\bibfnamefont {M.~A.~C.}\ \bibnamefont {Stuart}}, \
  and\ \bibinfo {author} {\bibfnamefont {M.}~\bibnamefont {Dijkstra}},\ }\href
  {\doibase 10.1038/ncomms3704} {\bibfield  {journal} {\bibinfo  {journal}
  {Nat. Com.}\ }\textbf {\bibinfo {volume} {4}},\ \bibinfo {pages} {2704}
  (\bibinfo {year} {2013})}\BibitemShut {NoStop}%
\bibitem [{\citenamefont {Berthier}(2014)}]{Berthier:14}%
  \BibitemOpen
  \bibfield  {author} {\bibinfo {author} {\bibfnamefont {L.}~\bibnamefont
  {Berthier}},\ }\href {\doibase 10.1103/PhysRevLett.112.220602} {\bibfield
  {journal} {\bibinfo  {journal} {Phys. Rev. Lett.}\ }\textbf {\bibinfo
  {volume} {112}},\ \bibinfo {pages} {220602} (\bibinfo {year}
  {2014})}\BibitemShut {NoStop}%
\bibitem [{\citenamefont {Levis}\ and\ \citenamefont
  {Berthier}(2014)}]{Levis:14}%
  \BibitemOpen
  \bibfield  {author} {\bibinfo {author} {\bibfnamefont {D.}~\bibnamefont
  {Levis}}\ and\ \bibinfo {author} {\bibfnamefont {L.}~\bibnamefont
  {Berthier}},\ }\href {\doibase 10.1103/PhysRevE.89.062301} {\bibfield
  {journal} {\bibinfo  {journal} {Phys. Rev. E}\ }\textbf {\bibinfo {volume}
  {89}},\ \bibinfo {pages} {062301} (\bibinfo {year} {2014})}\BibitemShut
  {NoStop}%
\bibitem [{\citenamefont {{Mandal}}\ \emph {et~al.}(2014)\citenamefont
  {{Mandal}}, \citenamefont {{Jyoti Bhuyan}}, \citenamefont {{Rao}},\ and\
  \citenamefont {{Dasgupta}}}]{Rao:14}%
  \BibitemOpen
  \bibfield  {author} {\bibinfo {author} {\bibfnamefont {R.}~\bibnamefont
  {{Mandal}}}, \bibinfo {author} {\bibfnamefont {P.}~\bibnamefont {{Jyoti
  Bhuyan}}}, \bibinfo {author} {\bibfnamefont {M.}~\bibnamefont {{Rao}}}, \
  and\ \bibinfo {author} {\bibfnamefont {C.}~\bibnamefont {{Dasgupta}}},\
  }\href@noop {} {\bibfield  {journal} {\bibinfo  {journal} {ArXiv e-prints}\ }
  (\bibinfo {year} {2014})},\ \Eprint {http://arxiv.org/abs/1412.1631}
  {arXiv:1412.1631} \BibitemShut {NoStop}%
\bibitem [{\citenamefont {Wysocki}\ \emph {et~al.}(2014)\citenamefont
  {Wysocki}, \citenamefont {Winkler},\ and\ \citenamefont
  {Gompper}}]{Gomper:14}%
  \BibitemOpen
  \bibfield  {author} {\bibinfo {author} {\bibfnamefont {A.}~\bibnamefont
  {Wysocki}}, \bibinfo {author} {\bibfnamefont {R.~G.}\ \bibnamefont
  {Winkler}}, \ and\ \bibinfo {author} {\bibfnamefont {G.}~\bibnamefont
  {Gompper}},\ }\href {\doibase 10.1209/0295-5075/105/48004} {\bibfield
  {journal} {\bibinfo  {journal} {EPL}\ }\textbf {\bibinfo {volume} {105}},\
  \bibinfo {pages} {48004} (\bibinfo {year} {2014})}\BibitemShut {NoStop}%
\bibitem [{\citenamefont {Berthier}\ and\ \citenamefont
  {Kurchan}(2013)}]{Kurchan:13}%
  \BibitemOpen
  \bibfield  {author} {\bibinfo {author} {\bibfnamefont {L.}~\bibnamefont
  {Berthier}}\ and\ \bibinfo {author} {\bibfnamefont {J.}~\bibnamefont
  {Kurchan}},\ }\href {\doibase 10.1038/nphys2592} {\bibfield  {journal}
  {\bibinfo  {journal} {Nat. Phys.}\ }\textbf {\bibinfo {volume} {9}},\
  \bibinfo {pages} {310} (\bibinfo {year} {2013})}\BibitemShut {NoStop}%
\bibitem [{\citenamefont {Szamel}\ \emph {et~al.}(2015)\citenamefont {Szamel},
  \citenamefont {Flenner},\ and\ \citenamefont {Berthier}}]{Szamel:15}%
  \BibitemOpen
  \bibfield  {author} {\bibinfo {author} {\bibfnamefont {G.}~\bibnamefont
  {Szamel}}, \bibinfo {author} {\bibfnamefont {E.}~\bibnamefont {Flenner}}, \
  and\ \bibinfo {author} {\bibfnamefont {L.}~\bibnamefont {Berthier}},\ }\href
  {\doibase 10.1103/PhysRevE.91.062304} {\bibfield  {journal} {\bibinfo
  {journal} {Phys. Rev. E}\ }\textbf {\bibinfo {volume} {91}},\ \bibinfo
  {pages} {062304} (\bibinfo {year} {2015})}\BibitemShut {NoStop}%
\bibitem [{\citenamefont {{Farage}}\ and\ \citenamefont
  {{Brader}}(2014)}]{Brader:14}%
  \BibitemOpen
  \bibfield  {author} {\bibinfo {author} {\bibfnamefont {T.~F.~F.}\
  \bibnamefont {{Farage}}}\ and\ \bibinfo {author} {\bibfnamefont {J.~M.}\
  \bibnamefont {{Brader}}},\ }\href@noop {} {\bibfield  {journal} {\bibinfo
  {journal} {ArXiv e-prints}\ } (\bibinfo {year} {2014})},\ \Eprint
  {http://arxiv.org/abs/1403.0928} {arXiv:1403.0928} \BibitemShut {NoStop}%
\bibitem [{\citenamefont {Berthier}\ and\ \citenamefont
  {Kob}(2007)}]{Berthier:07b}%
  \BibitemOpen
  \bibfield  {author} {\bibinfo {author} {\bibfnamefont {L.}~\bibnamefont
  {Berthier}}\ and\ \bibinfo {author} {\bibfnamefont {W.}~\bibnamefont {Kob}},\
  }\href {\doibase 10.1088/0953-8984/19/20/205130} {\bibfield  {journal}
  {\bibinfo  {journal} {J. Phys.: Cond. Mat.}\ }\textbf {\bibinfo {volume}
  {19}},\ \bibinfo {pages} {205130} (\bibinfo {year} {2007})}\BibitemShut
  {NoStop}%
\bibitem [{\citenamefont {Scalliet}\ \emph {et~al.}(2015)\citenamefont
  {Scalliet}, \citenamefont {Gnoli}, \citenamefont {Puglisi},\ and\
  \citenamefont {Vulpiani}}]{Scalliet:15}%
  \BibitemOpen
  \bibfield  {author} {\bibinfo {author} {\bibfnamefont {C.}~\bibnamefont
  {Scalliet}}, \bibinfo {author} {\bibfnamefont {A.}~\bibnamefont {Gnoli}},
  \bibinfo {author} {\bibfnamefont {A.}~\bibnamefont {Puglisi}}, \ and\
  \bibinfo {author} {\bibfnamefont {A.}~\bibnamefont {Vulpiani}},\ }\href
  {\doibase 10.1103/PhysRevLett.114.198001} {\bibfield  {journal} {\bibinfo
  {journal} {Phys. Rev. Lett.}\ }\textbf {\bibinfo {volume} {114}},\ \bibinfo
  {pages} {198001} (\bibinfo {year} {2015})}\BibitemShut {NoStop}%
\bibitem [{\citenamefont {Lasanta}\ and\ \citenamefont
  {Puglisi}(2015)}]{Lasanta:15}%
  \BibitemOpen
  \bibfield  {author} {\bibinfo {author} {\bibfnamefont {A.}~\bibnamefont
  {Lasanta}}\ and\ \bibinfo {author} {\bibfnamefont {A.}~\bibnamefont
  {Puglisi}},\ }\href {\doibase http://dx.doi.org/10.1063/1.4928456} {\bibfield
   {journal} {\bibinfo  {journal} {The Journal of Chemical Physics}\ }\textbf
  {\bibinfo {volume} {143}},\ \bibinfo {eid} {064511} (\bibinfo {year}
  {2015}),\ http://dx.doi.org/10.1063/1.4928456}\BibitemShut {NoStop}%
\bibitem [{\citenamefont {Kanazawa}\ \emph {et~al.}(2015)\citenamefont
  {Kanazawa}, \citenamefont {Sano}, \citenamefont {Sagawa},\ and\ \citenamefont
  {Hayakawa}}]{Kanazawa:15}%
  \BibitemOpen
  \bibfield  {author} {\bibinfo {author} {\bibfnamefont {K.}~\bibnamefont
  {Kanazawa}}, \bibinfo {author} {\bibfnamefont {T.~G.}\ \bibnamefont {Sano}},
  \bibinfo {author} {\bibfnamefont {T.}~\bibnamefont {Sagawa}}, \ and\ \bibinfo
  {author} {\bibfnamefont {H.}~\bibnamefont {Hayakawa}},\ }\href {\doibase
  10.1103/PhysRevLett.114.090601} {\bibfield  {journal} {\bibinfo  {journal}
  {Phys. Rev. Lett.}\ }\textbf {\bibinfo {volume} {114}},\ \bibinfo {pages}
  {090601} (\bibinfo {year} {2015})}\BibitemShut {NoStop}%
\bibitem [{\citenamefont {C\'aceres}\ and\ \citenamefont
  {Budini}(1997)}]{Budini}%
  \BibitemOpen
  \bibfield  {author} {\bibinfo {author} {\bibfnamefont {M.~O.}\ \bibnamefont
  {C\'aceres}}\ and\ \bibinfo {author} {\bibfnamefont {A.~A.}\ \bibnamefont
  {Budini}},\ }\href {\doibase 10.1088/0305-4470/30/24/009} {\bibfield
  {journal} {\bibinfo  {journal} {J. Phys. A}\ }\textbf {\bibinfo {volume}
  {30}},\ \bibinfo {pages} {8427} (\bibinfo {year} {1997})}\BibitemShut
  {NoStop}%
\bibitem [{\citenamefont {Baule}\ and\ \citenamefont
  {Cohen}(2009{\natexlab{a}})}]{Baule:09R}%
  \BibitemOpen
  \bibfield  {author} {\bibinfo {author} {\bibfnamefont {A.}~\bibnamefont
  {Baule}}\ and\ \bibinfo {author} {\bibfnamefont {E.~G.~D.}\ \bibnamefont
  {Cohen}},\ }\href {\doibase 10.1103/PhysRevE.79.030103} {\bibfield  {journal}
  {\bibinfo  {journal} {Phys. Rev. E}\ }\textbf {\bibinfo {volume} {79}},\
  \bibinfo {pages} {030103} (\bibinfo {year} {2009}{\natexlab{a}})}\BibitemShut
  {NoStop}%
\bibitem [{\citenamefont {Baule}\ and\ \citenamefont
  {Cohen}(2009{\natexlab{b}})}]{Baule:09}%
  \BibitemOpen
  \bibfield  {author} {\bibinfo {author} {\bibfnamefont {A.}~\bibnamefont
  {Baule}}\ and\ \bibinfo {author} {\bibfnamefont {E.~G.~D.}\ \bibnamefont
  {Cohen}},\ }\href {\doibase 10.1103/PhysRevE.80.011110} {\bibfield  {journal}
  {\bibinfo  {journal} {Phys. Rev. E}\ }\textbf {\bibinfo {volume} {80}},\
  \bibinfo {pages} {011110} (\bibinfo {year} {2009}{\natexlab{b}})}\BibitemShut
  {NoStop}%
\bibitem [{\citenamefont {Weeks}\ \emph {et~al.}(2007)\citenamefont {Weeks},
  \citenamefont {Crocker},\ and\ \citenamefont {Weitz}}]{Weeks:07}%
  \BibitemOpen
  \bibfield  {author} {\bibinfo {author} {\bibfnamefont {E.~R.}\ \bibnamefont
  {Weeks}}, \bibinfo {author} {\bibfnamefont {J.~C.}\ \bibnamefont {Crocker}},
  \ and\ \bibinfo {author} {\bibfnamefont {D.~A.}\ \bibnamefont {Weitz}},\
  }\href {\doibase 10.1088/0953-8984/19/20/205131} {\bibfield  {journal}
  {\bibinfo  {journal} {J. Phys.: Condens. Matter}\ }\textbf {\bibinfo {volume}
  {19}},\ \bibinfo {pages} {205131} (\bibinfo {year} {2007})}\BibitemShut
  {NoStop}%
\bibitem [{\citenamefont {{Franklin}}\ and\ \citenamefont
  {{Weeks}}(2014)}]{Weeks:15}%
  \BibitemOpen
  \bibfield  {author} {\bibinfo {author} {\bibfnamefont {S.~V.}\ \bibnamefont
  {{Franklin}}}\ and\ \bibinfo {author} {\bibfnamefont {E.~R.}\ \bibnamefont
  {{Weeks}}},\ }\href@noop {} {\bibfield  {journal} {\bibinfo  {journal} {ArXiv
  e-prints}\ } (\bibinfo {year} {2014})},\ \Eprint
  {http://arxiv.org/abs/1406.5782} {arXiv:1406.5782} \BibitemShut {NoStop}%
\bibitem [{\citenamefont {Kob}\ \emph {et~al.}(1997)\citenamefont {Kob},
  \citenamefont {Donati}, \citenamefont {Plimpton}, \citenamefont {Poole},\
  and\ \citenamefont {Glotzer}}]{Kob:97}%
  \BibitemOpen
  \bibfield  {author} {\bibinfo {author} {\bibfnamefont {W.}~\bibnamefont
  {Kob}}, \bibinfo {author} {\bibfnamefont {C.}~\bibnamefont {Donati}},
  \bibinfo {author} {\bibfnamefont {S.~J.}\ \bibnamefont {Plimpton}}, \bibinfo
  {author} {\bibfnamefont {P.~H.}\ \bibnamefont {Poole}}, \ and\ \bibinfo
  {author} {\bibfnamefont {S.~C.}\ \bibnamefont {Glotzer}},\ }\href {\doibase
  10.1103/PhysRevLett.79.2827} {\bibfield  {journal} {\bibinfo  {journal}
  {Phys. Rev. Lett.}\ }\textbf {\bibinfo {volume} {79}},\ \bibinfo {pages}
  {2827} (\bibinfo {year} {1997})}\BibitemShut {NoStop}%
\bibitem [{\citenamefont {Narumi}\ \emph {et~al.}(2011)\citenamefont {Narumi},
  \citenamefont {Franklin}, \citenamefont {Desmond}, \citenamefont {Tokuyama},\
  and\ \citenamefont {Weeks}}]{Weeks:11}%
  \BibitemOpen
  \bibfield  {author} {\bibinfo {author} {\bibfnamefont {T.}~\bibnamefont
  {Narumi}}, \bibinfo {author} {\bibfnamefont {S.~V.}\ \bibnamefont
  {Franklin}}, \bibinfo {author} {\bibfnamefont {K.~W.}\ \bibnamefont
  {Desmond}}, \bibinfo {author} {\bibfnamefont {M.}~\bibnamefont {Tokuyama}}, \
  and\ \bibinfo {author} {\bibfnamefont {E.~R.}\ \bibnamefont {Weeks}},\ }\href
  {\doibase 10.1039/C0SM00756K} {\bibfield  {journal} {\bibinfo  {journal}
  {Soft Matter}\ }\textbf {\bibinfo {volume} {7}},\ \bibinfo {pages} {1472}
  (\bibinfo {year} {2011})}\BibitemShut {NoStop}%
\bibitem [{\citenamefont {Charbonneau}\ \emph {et~al.}(2012)\citenamefont
  {Charbonneau}, \citenamefont {Ikeda}, \citenamefont {Parisi},\ and\
  \citenamefont {Zamponi}}]{Ikeda:12}%
  \BibitemOpen
  \bibfield  {author} {\bibinfo {author} {\bibfnamefont {P.}~\bibnamefont
  {Charbonneau}}, \bibinfo {author} {\bibfnamefont {A.}~\bibnamefont {Ikeda}},
  \bibinfo {author} {\bibfnamefont {G.}~\bibnamefont {Parisi}}, \ and\ \bibinfo
  {author} {\bibfnamefont {F.}~\bibnamefont {Zamponi}},\ }\href {\doibase
  10.1073/pnas.1211825109} {\bibfield  {journal} {\bibinfo  {journal} {Proc.
  Natl. Acad. Sci. USA}\ }\textbf {\bibinfo {volume} {109}},\ \bibinfo {pages}
  {13939} (\bibinfo {year} {2012})}\BibitemShut {NoStop}%
\bibitem [{\citenamefont {Vogel}\ and\ \citenamefont {Glotzer}(2004)}]{Vogel}%
  \BibitemOpen
  \bibfield  {author} {\bibinfo {author} {\bibfnamefont {M.}~\bibnamefont
  {Vogel}}\ and\ \bibinfo {author} {\bibfnamefont {S.~C.}\ \bibnamefont
  {Glotzer}},\ }\href {\doibase 10.1103/PhysRevE.70.061504} {\bibfield
  {journal} {\bibinfo  {journal} {Phys. Rev. E}\ }\textbf {\bibinfo {volume}
  {70}},\ \bibinfo {pages} {061504} (\bibinfo {year} {2004})}\BibitemShut
  {NoStop}%
\bibitem [{\citenamefont {Bacher}\ \emph {et~al.}(2013)\citenamefont {Bacher},
  \citenamefont {Schr{\o}der},\ and\ \citenamefont {Dyre}}]{Dyre}%
  \BibitemOpen
  \bibfield  {author} {\bibinfo {author} {\bibfnamefont {A.~K.}\ \bibnamefont
  {Bacher}}, \bibinfo {author} {\bibfnamefont {T.~B.}\ \bibnamefont
  {Schr{\o}der}}, \ and\ \bibinfo {author} {\bibfnamefont {J.~C.}\ \bibnamefont
  {Dyre}},\ }\href {\doibase 10.1038/ncomms6424} {\bibfield  {journal}
  {\bibinfo  {journal} {Nat. Com.}\ }\textbf {\bibinfo {volume} {5}},\ \bibinfo
  {pages} {5424} (\bibinfo {year} {2013})}\BibitemShut {NoStop}%
\bibitem [{\citenamefont {Bursac}\ \emph {et~al.}(2005)\citenamefont {Bursac},
  \citenamefont {Lenormand}, \citenamefont {Oliver}, \citenamefont {Weitz},
  \citenamefont {Viasnoff}, \citenamefont {Butler},\ and\ \citenamefont
  {Fredberg}}]{Bursac}%
  \BibitemOpen
  \bibfield  {author} {\bibinfo {author} {\bibfnamefont {P.}~\bibnamefont
  {Bursac}}, \bibinfo {author} {\bibfnamefont {G.}~\bibnamefont {Lenormand}},
  \bibinfo {author} {\bibfnamefont {B.~F.~M.}\ \bibnamefont {Oliver}}, \bibinfo
  {author} {\bibfnamefont {D.~A.}\ \bibnamefont {Weitz}}, \bibinfo {author}
  {\bibfnamefont {V.}~\bibnamefont {Viasnoff}}, \bibinfo {author}
  {\bibfnamefont {J.~P.}\ \bibnamefont {Butler}}, \ and\ \bibinfo {author}
  {\bibfnamefont {J.~J.}\ \bibnamefont {Fredberg}},\ }\href {\doibase
  10.1038/nmat1404} {\bibfield  {journal} {\bibinfo  {journal} {Nat. Mat.}\
  }\textbf {\bibinfo {volume} {4}},\ \bibinfo {pages} {557} (\bibinfo {year}
  {2005})}\BibitemShut {NoStop}%
\bibitem [{\citenamefont {Angelini}\ \emph {et~al.}(2011)\citenamefont
  {Angelini}, \citenamefont {Hannezo}, \citenamefont {Trepat}, \citenamefont
  {Marquez}, \citenamefont {Fredberg},\ and\ \citenamefont
  {Weitz}}]{Hannezo:11}%
  \BibitemOpen
  \bibfield  {author} {\bibinfo {author} {\bibfnamefont {T.~E.}\ \bibnamefont
  {Angelini}}, \bibinfo {author} {\bibfnamefont {E.}~\bibnamefont {Hannezo}},
  \bibinfo {author} {\bibfnamefont {X.}~\bibnamefont {Trepat}}, \bibinfo
  {author} {\bibfnamefont {M.}~\bibnamefont {Marquez}}, \bibinfo {author}
  {\bibfnamefont {J.~J.}\ \bibnamefont {Fredberg}}, \ and\ \bibinfo {author}
  {\bibfnamefont {D.~A.}\ \bibnamefont {Weitz}},\ }\href {\doibase
  10.1073/pnas.1010059108} {\bibfield  {journal} {\bibinfo  {journal} {Proc.
  Natl. Acad. Sci. USA}\ }\textbf {\bibinfo {volume} {108}},\ \bibinfo {pages}
  {4714} (\bibinfo {year} {2011})}\BibitemShut {NoStop}%
\bibitem [{\citenamefont {Brangwynne}\ \emph {et~al.}(2009)\citenamefont
  {Brangwynne}, \citenamefont {Koenderink}, \citenamefont {MacKintosh},\ and\
  \citenamefont {Weitz}}]{Weitz:09}%
  \BibitemOpen
  \bibfield  {author} {\bibinfo {author} {\bibfnamefont {C.~P.}\ \bibnamefont
  {Brangwynne}}, \bibinfo {author} {\bibfnamefont {G.~H.}\ \bibnamefont
  {Koenderink}}, \bibinfo {author} {\bibfnamefont {F.~C.}\ \bibnamefont
  {MacKintosh}}, \ and\ \bibinfo {author} {\bibfnamefont {D.~A.}\ \bibnamefont
  {Weitz}},\ }\href {\doibase 10.1016/j.tcb.2009.04.004} {\bibfield  {journal}
  {\bibinfo  {journal} {Trends Cell Biol.}\ }\textbf {\bibinfo {volume} {19}},\
  \bibinfo {pages} {423 } (\bibinfo {year} {2009})}\BibitemShut {NoStop}%
\bibitem [{\citenamefont {Ahmed}\ \emph {et~al.}(2015)\citenamefont {Ahmed},
  \citenamefont {Fodor},\ and\ \citenamefont {Betz}}]{Ahmed:15}%
  \BibitemOpen
  \bibfield  {author} {\bibinfo {author} {\bibfnamefont {W.~W.}\ \bibnamefont
  {Ahmed}}, \bibinfo {author} {\bibfnamefont {{\'E}.}~\bibnamefont {Fodor}}, \
  and\ \bibinfo {author} {\bibfnamefont {T.}~\bibnamefont {Betz}},\ }\href
  {\doibase 10.1016/j.bbamcr.2015.05.022} {\bibfield  {journal} {\bibinfo
  {journal} {BBA - Mol. Cell Res.}\ }\textbf {\bibinfo {volume} {1853}},\
  \bibinfo {pages} {3083} (\bibinfo {year} {2015})}\BibitemShut {NoStop}%
\bibitem [{\citenamefont {Zaid}\ \emph {et~al.}(2015)\citenamefont {Zaid},
  \citenamefont {Ayade},\ and\ \citenamefont {Mizuno}}]{Mizuno:14a}%
  \BibitemOpen
  \bibfield  {author} {\bibinfo {author} {\bibfnamefont {I.}~\bibnamefont
  {Zaid}}, \bibinfo {author} {\bibfnamefont {H.~L.}\ \bibnamefont {Ayade}}, \
  and\ \bibinfo {author} {\bibfnamefont {D.}~\bibnamefont {Mizuno}},\ }\href
  {\doibase 10.1016/j.bpj.2013.11.973} {\bibfield  {journal} {\bibinfo
  {journal} {Biophys. J.}\ }\textbf {\bibinfo {volume} {106}} (\bibinfo {year}
  {2015}),\ 10.1016/j.bpj.2013.11.973}\BibitemShut {NoStop}%
\bibitem [{\citenamefont {Aridome}\ \emph {et~al.}(2015)\citenamefont
  {Aridome}, \citenamefont {Kurihara}, \citenamefont {Ayade}, \citenamefont
  {Zaid},\ and\ \citenamefont {Mizuno}}]{Mizuno:14b}%
  \BibitemOpen
  \bibfield  {author} {\bibinfo {author} {\bibfnamefont {M.}~\bibnamefont
  {Aridome}}, \bibinfo {author} {\bibfnamefont {T.}~\bibnamefont {Kurihara}},
  \bibinfo {author} {\bibfnamefont {H.~L.}\ \bibnamefont {Ayade}}, \bibinfo
  {author} {\bibfnamefont {I.}~\bibnamefont {Zaid}}, \ and\ \bibinfo {author}
  {\bibfnamefont {D.}~\bibnamefont {Mizuno}},\ }\href {\doibase
  10.1016/j.bpj.2013.11.3218} {\bibfield  {journal} {\bibinfo  {journal}
  {Biophys. J.}\ }\textbf {\bibinfo {volume} {106}} (\bibinfo {year} {2015}),\
  10.1016/j.bpj.2013.11.3218}\BibitemShut {NoStop}%
\bibitem [{\citenamefont {Fodor}\ \emph {et~al.}(2014)\citenamefont {Fodor},
  \citenamefont {Kanazawa}, \citenamefont {Hayakawa}, \citenamefont {Visco},\
  and\ \citenamefont {van Wijland}}]{Kanazawa:14}%
  \BibitemOpen
  \bibfield  {author} {\bibinfo {author} {\bibfnamefont {{\'E}.}~\bibnamefont
  {Fodor}}, \bibinfo {author} {\bibfnamefont {K.}~\bibnamefont {Kanazawa}},
  \bibinfo {author} {\bibfnamefont {H.}~\bibnamefont {Hayakawa}}, \bibinfo
  {author} {\bibfnamefont {P.}~\bibnamefont {Visco}}, \ and\ \bibinfo {author}
  {\bibfnamefont {F.}~\bibnamefont {van Wijland}},\ }\href {\doibase
  10.1103/PhysRevE.90.042724} {\bibfield  {journal} {\bibinfo  {journal} {Phys.
  Rev. E}\ }\textbf {\bibinfo {volume} {90}},\ \bibinfo {pages} {042724}
  (\bibinfo {year} {2014})}\BibitemShut {NoStop}%
\bibitem [{\citenamefont {Fodor}\ \emph {et~al.}(2015)\citenamefont {Fodor},
  \citenamefont {Guo}, \citenamefont {Gov}, \citenamefont {Visco},
  \citenamefont {Weitz},\ and\ \citenamefont {van Wijland}}]{Visco:15}%
  \BibitemOpen
  \bibfield  {author} {\bibinfo {author} {\bibfnamefont {{\'E}.}~\bibnamefont
  {Fodor}}, \bibinfo {author} {\bibfnamefont {M.}~\bibnamefont {Guo}}, \bibinfo
  {author} {\bibfnamefont {N.~S.}\ \bibnamefont {Gov}}, \bibinfo {author}
  {\bibfnamefont {P.}~\bibnamefont {Visco}}, \bibinfo {author} {\bibfnamefont
  {D.~A.}\ \bibnamefont {Weitz}}, \ and\ \bibinfo {author} {\bibfnamefont
  {F.}~\bibnamefont {van Wijland}},\ }\href {\doibase
  10.1209/0295-5075/110/48005} {\bibfield  {journal} {\bibinfo  {journal}
  {EPL}\ }\textbf {\bibinfo {volume} {110}},\ \bibinfo {pages} {48005}
  (\bibinfo {year} {2015})}\BibitemShut {NoStop}%
\bibitem [{\citenamefont {Ben-Isaac}\ \emph {et~al.}(2015)\citenamefont
  {Ben-Isaac}, \citenamefont {Fodor}, \citenamefont {Visco}, \citenamefont {van
  Wijland},\ and\ \citenamefont {Gov}}]{Gov:15}%
  \BibitemOpen
  \bibfield  {author} {\bibinfo {author} {\bibfnamefont {E.}~\bibnamefont
  {Ben-Isaac}}, \bibinfo {author} {\bibfnamefont {{\'E}.}~\bibnamefont
  {Fodor}}, \bibinfo {author} {\bibfnamefont {P.}~\bibnamefont {Visco}},
  \bibinfo {author} {\bibfnamefont {F.}~\bibnamefont {van Wijland}}, \ and\
  \bibinfo {author} {\bibfnamefont {N.~S.}\ \bibnamefont {Gov}},\ }\href
  {\doibase 10.1103/PhysRevE.92.012716} {\bibfield  {journal} {\bibinfo
  {journal} {Phys. Rev. E}\ }\textbf {\bibinfo {volume} {92}},\ \bibinfo
  {pages} {012716} (\bibinfo {year} {2015})}\BibitemShut {NoStop}%
\bibitem [{\citenamefont {{Fodor}}\ \emph
  {et~al.}(2015{\natexlab{a}})\citenamefont {{Fodor}}, \citenamefont
  {{Mehandia}}, \citenamefont {{Comelles}}, \citenamefont {{Thiagarajan}},
  \citenamefont {{Gov}}, \citenamefont {{Visco}}, \citenamefont {{van
  Wijland}},\ and\ \citenamefont {{Riveline}}}]{Riveline:15}%
  \BibitemOpen
  \bibfield  {author} {\bibinfo {author} {\bibfnamefont {{\'E}.}~\bibnamefont
  {{Fodor}}}, \bibinfo {author} {\bibfnamefont {V.}~\bibnamefont {{Mehandia}}},
  \bibinfo {author} {\bibfnamefont {J.}~\bibnamefont {{Comelles}}}, \bibinfo
  {author} {\bibfnamefont {R.}~\bibnamefont {{Thiagarajan}}}, \bibinfo {author}
  {\bibfnamefont {N.~S.}\ \bibnamefont {{Gov}}}, \bibinfo {author}
  {\bibfnamefont {P.}~\bibnamefont {{Visco}}}, \bibinfo {author} {\bibfnamefont
  {F.}~\bibnamefont {{van Wijland}}}, \ and\ \bibinfo {author} {\bibfnamefont
  {D.}~\bibnamefont {{Riveline}}},\ }\href@noop {} {\bibfield  {journal}
  {\bibinfo  {journal} {ArXiv e-prints}\ } (\bibinfo {year}
  {2015}{\natexlab{a}})},\ \Eprint {http://arxiv.org/abs/1512.01476}
  {arXiv:1512.01476} \BibitemShut {NoStop}%
\bibitem [{\citenamefont {{Ahmed}}\ \emph {et~al.}(2015)\citenamefont
  {{Ahmed}}, \citenamefont {{Fodor}}, \citenamefont {{Almonacid}},
  \citenamefont {{Bussonnier}}, \citenamefont {{Verlhac}}, \citenamefont
  {{Gov}}, \citenamefont {{Visco}}, \citenamefont {{van Wijland}},\ and\
  \citenamefont {{Betz}}}]{Ahmed:15b}%
  \BibitemOpen
  \bibfield  {author} {\bibinfo {author} {\bibfnamefont {W.~W.}\ \bibnamefont
  {{Ahmed}}}, \bibinfo {author} {\bibfnamefont {E.}~\bibnamefont {{Fodor}}},
  \bibinfo {author} {\bibfnamefont {M.}~\bibnamefont {{Almonacid}}}, \bibinfo
  {author} {\bibfnamefont {M.}~\bibnamefont {{Bussonnier}}}, \bibinfo {author}
  {\bibfnamefont {M.-H.}\ \bibnamefont {{Verlhac}}}, \bibinfo {author}
  {\bibfnamefont {N.~S.}\ \bibnamefont {{Gov}}}, \bibinfo {author}
  {\bibfnamefont {P.}~\bibnamefont {{Visco}}}, \bibinfo {author} {\bibfnamefont
  {F.}~\bibnamefont {{van Wijland}}}, \ and\ \bibinfo {author} {\bibfnamefont
  {T.}~\bibnamefont {{Betz}}},\ }\href@noop {} {\bibfield  {journal} {\bibinfo
  {journal} {ArXiv e-prints}\ } (\bibinfo {year} {2015})},\ \Eprint
  {http://arxiv.org/abs/1510.08299} {arXiv:1510.08299} \BibitemShut {NoStop}%
\bibitem [{\citenamefont {{Fodor}}\ \emph
  {et~al.}(2015{\natexlab{b}})\citenamefont {{Fodor}}, \citenamefont {{Ahmed}},
  \citenamefont {{Almonacid}}, \citenamefont {{Bussonnier}}, \citenamefont
  {{Gov}}, \citenamefont {{Verlhac}}, \citenamefont {{Betz}}, \citenamefont
  {{Visco}},\ and\ \citenamefont {{van Wijland}}}]{Ahmed:15c}%
  \BibitemOpen
  \bibfield  {author} {\bibinfo {author} {\bibfnamefont {{\'E}.}~\bibnamefont
  {{Fodor}}}, \bibinfo {author} {\bibfnamefont {W.~W.}\ \bibnamefont
  {{Ahmed}}}, \bibinfo {author} {\bibfnamefont {M.}~\bibnamefont
  {{Almonacid}}}, \bibinfo {author} {\bibfnamefont {M.}~\bibnamefont
  {{Bussonnier}}}, \bibinfo {author} {\bibfnamefont {N.~S.}\ \bibnamefont
  {{Gov}}}, \bibinfo {author} {\bibfnamefont {M.-H.}\ \bibnamefont
  {{Verlhac}}}, \bibinfo {author} {\bibfnamefont {T.}~\bibnamefont {{Betz}}},
  \bibinfo {author} {\bibfnamefont {P.}~\bibnamefont {{Visco}}}, \ and\
  \bibinfo {author} {\bibfnamefont {F.}~\bibnamefont {{van Wijland}}},\
  }\href@noop {} {\bibfield  {journal} {\bibinfo  {journal} {ArXiv e-prints}\ }
  (\bibinfo {year} {2015}{\natexlab{b}})},\ \Eprint
  {http://arxiv.org/abs/1511.00921} {arXiv:1511.00921} \BibitemShut {NoStop}%
\bibitem [{\citenamefont {Gal}\ \emph {et~al.}(2013)\citenamefont {Gal},
  \citenamefont {Lechtman-Goldstein},\ and\ \citenamefont {Weihs}}]{Weihs}%
  \BibitemOpen
  \bibfield  {author} {\bibinfo {author} {\bibfnamefont {N.}~\bibnamefont
  {Gal}}, \bibinfo {author} {\bibfnamefont {D.}~\bibnamefont
  {Lechtman-Goldstein}}, \ and\ \bibinfo {author} {\bibfnamefont
  {D.}~\bibnamefont {Weihs}},\ }\href {\doibase 10.1007/s00397-013-0694-6}
  {\bibfield  {journal} {\bibinfo  {journal} {Rheol. Acta}\ }\textbf {\bibinfo
  {volume} {52}},\ \bibinfo {pages} {425} (\bibinfo {year} {2013})}\BibitemShut
  {NoStop}%
\bibitem [{\citenamefont {Wang}\ \emph {et~al.}(2009)\citenamefont {Wang},
  \citenamefont {Anthony}, \citenamefont {Bae},\ and\ \citenamefont
  {Granick}}]{Granick}%
  \BibitemOpen
  \bibfield  {author} {\bibinfo {author} {\bibfnamefont {B.}~\bibnamefont
  {Wang}}, \bibinfo {author} {\bibfnamefont {S.~M.}\ \bibnamefont {Anthony}},
  \bibinfo {author} {\bibfnamefont {S.~C.}\ \bibnamefont {Bae}}, \ and\
  \bibinfo {author} {\bibfnamefont {S.}~\bibnamefont {Granick}},\ }\href
  {\doibase 10.1073/pnas.0903554106} {\bibfield  {journal} {\bibinfo  {journal}
  {Proc. Natl. Acad. Sci. USA}\ }\textbf {\bibinfo {volume} {106}},\ \bibinfo
  {pages} {15160} (\bibinfo {year} {2009})}\BibitemShut {NoStop}%
\end{thebibliography}%

\end{document}